\documentstyle[12pt,epsf]{article}

\newcommand{\bmat}{\left(\begin{array}}
\newcommand{\emat}{\end{array}\right)}
\def\NPB#1#2#3{Nucl. Phys. B{#1} (19#2) #3}
\def\PLB#1#2#3{Phys. Lett. B{#1} (19#2) #3}

\def\PRD#1#2#3{Phys. Rev. D{#1} (19#2) #3}
\def\PRL#1#2#3{Phys. Rev. Lett. {#1} (19#2) #3}

\def\MODA#1#2#3{Mod. Phys. Lett.  A{#1} (19#2) #3}

\def\yzero{\smash{\hbox{$y\kern-4pt\raise1pt\hbox{${}^\circ$}$}}}

\def\-{\hphantom{-}}
\def\ov{\overline}
\def\s2{\frac{1}{\sqrt2}}

\def\beq{\begin{equation}}
\def\eeq{\end{equation}}
\def\beqa{\begin{eqnarray}}
\def\eeqa{\end{eqnarray}}
\def\tr{{\rm tr \,}}
\def\Tr{{\rm Tr \,}}
\def\diag{{\rm diag \,}}

\def\IF{\relax{\rm I\kern-.18em F}}
\def\II{\relax{\rm I\kern-.18em I}}
\def\IP{\relax{\rm I\kern-.18em P}}

\def\cc{{\cal C}}
\def\ck{{\cal K}}

\def\cm{{\cal M}}
\def\NN{{\cal N}}

\def\Dsl{\,\raise.15ex\hbox{/}\mkern-13.5mu D} 

\def\IC{\bf C}
\def\IZ{\bf Z}
\def\z2z2{$\IC^3/(\IZ_2\times\IZ_2)$}

\def\negp{{O$6\,'\,^{-}$-O$6\,'\,^{-}$}}
\def\posp{{O$6\,'\,^{+}$-O$6\,'\,^{+}$}}
\def\posnegp{{O$6\,'\,^{+}$-O$6\,'\,^{-}$}}
\def\negposp{{O$6\,'\,^{-}$-O$6\,'\,^{+}$}}

\newcommand{\drawsquare}[2]{\hbox{%
\rule{#2pt}{#1pt}\hskip-#2pt
\rule{#1pt}{#2pt}\hskip-#1pt
\rule[#1pt]{#1pt}{#2pt}}\rule[#1pt]{#2pt}{#2pt}\hskip-#2pt
\rule{#2pt}{#1pt}}

\newcommand{\fund}{\raisebox{-.5pt}{\drawsquare{6.5}{0.4}}}
\newcommand{\Ysymm}{\raisebox{-.5pt}{\drawsquare{6.5}{0.4}}\hskip-0.4pt%
        \raisebox{-.5pt}{\drawsquare{6.5}{0.4}}}
\newcommand{\Yasymm}{\raisebox{-3.5pt}{\drawsquare{6.5}{0.4}}\hskip-6.9pt%
        \raisebox{3pt}{\drawsquare{6.5}{0.4}}}
\newcommand{\antifund}{\overline{\fund}}
\newcommand{\bYasymm}{\overline{\Yasymm}}

\topmargin
-1.5cm
\textwidth
15.5cm
\textheight
24cm
\oddsidemargin
0.7cm
\evensidemargin
1.2cm

\begin{document}

\makeatletter
\@addtoreset{equation}{section}
\makeatother
\renewcommand{\theequation}{\thesection.\arabic{equation}}
\pagestyle{empty}
\rightline{FTUAM-99/21}
\rightline{IASSNS-HEP-99/61}
\rightline{IFT-UAM/CSIC-99-25}

\rightline{\tt hep-th/9907074}
\vspace{0.5cm}
\begin{center}
\LARGE{$\NN=1$ Type IIA brane configurations, \\
Chirality and T-duality\\[10mm]}
\large{
J.~Park$^{\dag\,}$\footnote{\texttt{jaemo@ias.edu}}, R. 
Rabad\'an$^{\S\,}$\footnote{\texttt{rabadan@delta.ft.uam.es}} and 
A.~M.~Uranga$^{\dag\,}$\footnote{\texttt{uranga@ias.edu}} \\[2mm]} 
\small{
$^\dag$ {\em School of Natural Sciences, Institute for Advanced Study,}\\ 
{\em Olden Lane, Princeton NJ 08540, USA.} \\[4mm]
$^\S$ {\em Departamento. de F\'{\i}sica Te\'orica C-XI
and Instituto de F\'{\i}sica Te\'orica C-XVI,}\\
{\em Universidad Aut\'onoma de Madrid, Cantoblanco, 28049 Madrid, 
Spain.}\\[4mm]
}
\small{\bf Abstract} \\[7mm]
\end{center}

\begin{center}
\begin{minipage}[h]{14.0cm}

{\small We consider four-dimensional $\NN=1$ field theories realized by
type IIA brane configurations of NS-branes and D4-branes, in the presence 
of orientifold six-planes and D6-branes. These configurations are known to 
present interesting effects associated to the appearance of chiral symmetries 
and chiral matter in the four-dimensional field theory. We center on models 
with one compact direction (elliptic models) and show that, under T-duality, 
the configurations are mapped to a set of type IIB D3-branes probing $\NN=1$ 
orientifolds of $\IC^2/\IZ_N$ singularities. We explicitly construct these 
orientifolds, and show the field theories on the D3-brane probes indeed 
reproduces the field theories constructed using the IIA brane 
configurations. This T-duality map allows to understand the type IIB 
realization of several exotic brane dynamics effects on the type IIA side: 
Flavour doubling, the splitting of D6-branes and O6-planes in crossing a 
NS-brane and the effect of a non-zero type IIA cosmological constant
turn out to have surprisingly standard type IIB counterparts.
}

\end{minipage}
\end{center}

\newpage
\setcounter{page}{1}
\pagestyle{plain}
\renewcommand{\thefootnote}{\arabic{footnote}}
\setcounter{footnote}{0}

\section{Introduction}

One of the most interesting features of four-dimensional field theories that 
has been successfully reproduced in their realization in string theory is the 
appearance of chiral matter and chiral symmetries. The first realizations of 
$\NN=1$ supersymmetric gauge field theories in terms of brane configurations, 
in the spirit of \cite{hw}, involved a set of type IIA D4-branes (along the 
directions 01236), relatively rotated NS-fivebranes (along 012345 or 012389), 
and D6-branes (along 0123789) \cite{egk} (see review for further references). 
However, given their close relation with the $\NN=2$ models in \cite{wit4d} 
{\em via} a rotation of NS-branes \cite{barbon} which makes the adjoint matter
massive, they have vector-like matter content. Also, the chiral symmetries 
present in the field theory for massless flavours are not manifest in the 
brane realization.

The second of these issues was further explored in \cite{hanbro}, where it 
was noticed that the fundamental flavours arising from the D6-branes have 
generically quartic superpotential interactions, mediated by the massive 
adjoints, which only preserve the diagonal vector-like subgroup of the chiral 
symmetry. This superpotential can however be varied by rotating the D6-branes 
in the 45-89 plane, and vanishes for D6-branes oriented along 0123457 
(D6$'$-branes) \footnote{Further support for these superpotentials was 
provided in \cite{ahan}.}. The chiral symmetries in the field theory were 
argued to be manifest in the brane configuration when the D6$'$ branes are 
located on top of the NS-branes and split in half D6$'$-branes, of 
semi-infinite extent in the direction 7. The independent gauge symmetries on 
these `upper ' and `lower' half D6$'$-branes correspond to the independent 
chiral rotations of the matter multiplets, leading to the conclusion that one 
half D6$'$-brane, ending on a NS-brane with D4-branes suspended on it, leads 
to one chiral fundamental multiplet in the four-dimensional field theory 
\footnote{This proposal implies that, in gauge theories with several gauge 
factors, a single half D6$'$-brane ending on a NS-brane provides {\em two} 
fundamental chiral multiplets, one for each gauge factor arising from 
D4-branes ending on the NS-brane. This phenomenon is known as `flavour 
doubling' \cite{bhkl}.}. Unfortunately, the inability of the NS-branes to 
carry away the RR charge of the half D6$'$-branes requires the number of 
upper and lower half-branes to be identical, implying vector-like matter 
contents.

A further step was taken in \cite{hzd8}, where it was pointed out that the
presence of a non-zero type IIA cosmological constant (so that the 
configuration is actually embedded in massive type IIA theory) forces 
a mismatch between the number of upper and lower half D6$'$-branes (equal 
to the value of the cosmological constant in appropriate units), which 
allows to obtain chiral matter localized at these intersections. In this 
setup quite exotic phase transitions were shown to occur, in which the number 
of chiral multiplets changes due to a change in the IIA cosmological 
constant. The latter is achieved by introducing D8-branes (along 012345689), 
which behave as domain walls across which the cosmological constant 
changes, and crossing them through the configuration. In these initial models, 
however, overall conservation of RR charge implied the complete matter content 
must be vector-like, even though fundamental and anti-fundamental chiral 
multiplets are generically localized at spatially different positions in the 
direction 6. From the field theory point of view the vector-like matter 
content is merely consequence of cancellation of gauge anomalies, since the 
configuration only allow to obtain fundamental and anti-fundamental 
representations.
 
The final ingredient to achieve chiral matter contents was the introduction 
of orientifold planes parallel to the D6$'$-branes (O6$'$-planes), since 
they allow the appearance of two-index tensor representations. Specifically, 
when an O6$'$-plane sits on top of a NS-brane, its two halves have opposite 
RR charge \cite{ejs}. Due to the different orientifold projection in the upper 
and lower halves in $x^7$, there appears chiral matter multiplets transforming 
in the $\Ysymm + \bYasymm$ representation \cite{lll,egkt,bhkl}. Finally, 
overall conservation of RR charge (or anomaly cancellation in the field 
theory) implies the existence of additional half D6$'$-branes producing eight 
anti-fundamental chiral flavours \footnote{This chiral matter content 
has been studied in \cite{ils} from the field theory point of view.}, 
localized either at the same position in 
the direction 6 (for zero IIA cosmological constant) or elsewhere (for 
non-zero cosmological constant). This type of dynamics of D6-branes and 
O6-planes has also appeared in the context of type IIA brane configurations 
realizing six-dimensional supersymmetric field theories \cite{bk1,bk2,hz}.

Despite this success, which allows to realize a chiral gauge field theory 
in string theory, and to analyze its M-theory lift \cite{park,lll2}, the 
type IIA approach is extremely rigid and has not allowed to construct 
further examples of chiral theories \footnote{In \cite{lpt} the 
introduction of orbifold (and orientifold) singularities in the type IIA 
setup allowed to construct further chiral models. They can be regarded as 
T-duals of the brane box models mentioned below.}. Completely different 
setups, like type IIB brane box models \cite{hzbb}, and their T-dual 
realization in terms of D3-brane probes at threefold singularities 
\cite{d3probe, nonab, coni} (and orientifolds \footnote{Strings in 
orientifolds backgrounds were first introduced in \cite{sag1,dlp,horava,bs}} 
thereof \cite{d3orient}) have proved more efficient in producing generically 
chiral field theories.

\medskip

Two conclusions can be drawn from this brief review. The first is the 
surprising amount, diversity and apparent `exotism' of the non-trivial 
effects required in the type IIA setup to reproduce models with chiral 
symmetries and chiral matter content. The second is the apparent extreme 
isolation of the chiral gauge theory constructed in the IIA setup using 
the change of sign of the O6$'$-plane. It seems to be rather unrelated to 
other configurations in string theory realizing four-dimensional chiral 
gauge theories.

In this paper we provide some insight into these issues. We consider type 
IIA configurations of NS-branes (along 012345), D4-branes (along 01236), 
and D6$'$-branes and O6$'$-planes (along 0123457), with the direction 6 
compactified on a circle. These can be regarded as the $\NN=2$ elliptic 
models of \cite{wit4d} in the presence of O6$'$-planes which break the 
supersymmetry to $\NN=1$. We study the resulting type IIB string theory
configurations obtained upon T-dualizing along this compact direction. 
They correspond to $\NN=1$ orientifolds of a system of D3-branes at a 
$\IC^2/\IZ_N$ singularity. We explicitly classify and construct these 
orientifolds, and show they indeed realize the field theories obtained 
from the type IIA configurations. The matching of both types of models 
allows to find the type IIB counterparts of the non-trivial brane dynamics 
effects present in the IIA models. The correct chiral matter content and 
pattern of chiral symmetries are reproduced by the completely standard 
rules of IIB orientifold construction. We list here the main results:
\begin{itemize}
\item We find that half D6$'$-branes map to whole D7-branes. Concretely, 
upper and lower half D6$'$-branes map to D7-branes along either of the two 
complex planes in $\IC^2/\IZ_N$. The chiral symmetries in the field theory 
(gauge symmetries on the half D6$'$-branes) appear as gauge symmetries on the 
D7-branes  \footnote{This facts have already been observed in \cite{kg} in 
a related context.}.

\item The phenomenon of flavour doubling is reproduced by the fact that {\em 
one} D7-brane produces {\em two} fundamental chiral flavours arising from the 
3-7 and 7-3 sectors \footnote{In orientifold models, where both sectors are 
related by the orientifold projection, the {\em two} fundamental flavours 
still arise, in a slightly different fashion.}. 

\item The matter content $\Ysymm+\bYasymm+8\antifund$ in the chiral 
theories is correctly reproduced in the IIB setup. The IIB orientifold 
also encodes the structures of O6$'$-planes in the upper and lower halves 
in $x^7$, in terms of the orientifold action on the two complex planes in 
$\IC^2/\IZ_N$. The type IIB orientifold configurations are, in 
contrast with the type IIA models, quite smooth and do not present abrupt 
changes in the orientifold type.

\item Cancellation of anomalies in the field theory (conservation of RR 
charge in the IIA configuration) corresponds to cancellation of tadpoles 
for twisted RR fields in the IIB model.

\item We also find the type IIB T-dual of the type IIA cosmological 
constant. The transitions changing the number of field theory flavours are 
interpreted as a brane creation process in the type IIB side.
\end{itemize}

Thus, several rather `exotic' effects in the type IIA brane configurations 
are reproduced by completely standard rules in the context of type IIB 
orientifold construction. This also sheds some light on the issue of the 
isolation of these chiral field theories. In the type IIB realization as 
D3-branes probing orientifold singularities, these models do not differ in 
any essential way from the constructions yielding  generically chiral field 
theories (for instance D3-branes probes at $\IC^3/\Gamma$ threefold orbifold 
singularities and orientifolds thereof). This point becomes even clearer by 
noticing that both types of models are actually continuously connected 
\cite{pru}: Some configurations of D3-branes at orientifolds of $\IC^3/\Gamma$ 
reduce, after suitable partial resolutions of the singularity, to the 
orientifolds of $\IC^2/\IZ_N$ appearing in this paper; from the field theory 
point of view this is realized a suitable baryonic Higgs breaking that 
relates both field theories. Therefore, our realization of these models in the 
singularity picture points towards a more unified description of all 
string theory configurations producing four-dimensional chiral field theories.

A last advantage of having a realization of these gauge theories in terms 
of D3-brane probes at singularities is that it is well-suited for the 
study of the field theory in the large $N$ limit through the AdS/CFT 
correspondence \cite{revads}.

The paper is organized as follows. In Section~2 we construct type IIA 
configurations of NS-branes, D4-branes and D6$'$-branes, with one compact 
direction but without orientifold planes. Even though the fields can be 
arranged in $\NN=2$ multiplets, the superpotential interactions only preserve 
$\NN=1$ supersymmetry. We consider the case of D6$'$-branes sitting on top of 
the NS-branes, so that these configurations are the simplest models presenting 
the phenomena of manifest chiral symmetry, flavour doubling, localization of 
chiral matter, and non-trivial effect of the IIA cosmological constant. We 
construct the type IIB T-dual realization, in terms of D3-branes at a 
$\IC^2/\IZ_N$ singularity in the presence of a suitable set of D7-branes, 
and discuss the IIB counterparts of the above mentioned effects.

In Section~3 we consider the introduction of O6$'$-planes and present a 
complete classification of models with one compact direction. These illustrate 
other effects like the appearance of chiral matter content, due to the change 
of sign of the O6$'$-plane when crossing the NS-brane. In Section~4 we 
construct the T-dual type IIB version of these models. We classify the 
possible choices of Chan-Paton embeddings on D3-branes and find that they 
reproduce the field theories constructed in the IIA framework, including 
the models with chiral matter content. We discuss how the change of sign 
of the IIA O6$'$-planes is encoded in the IIB picture. 

In Section~5 we consider type IIB configurations with more general 
structure of D7-branes, and construct the corresponding IIA models, which 
include additional D6$'$-branes. We compute the tadpole cancellation 
conditions on the IIB orientifolds and find they are equivalent to the 
cancellation of gauge anomalies in the field theory, and to conservation 
of RR charge in the T-dual IIA brane configuration. 
We conclude with Section~6, which contains some final comments.

\medskip

\section{Flavour doubling and orbifold models}

Before entering the discussion of the brane configurations with orientifold 
projection in the coming sections, it is convenient to study some properties 
of $\NN=1$ brane configuration that appear even before the introduction of 
orientifolds. The IIA configurations we study in this section constitute an
interesting modification of the $\NN=2$ elliptic models in 
\cite{wit4d}, consisting on the introduction of D6$'$-branes. These break 
the supersymmetry to $\NN=1$ in the four-dimensional field theory. The 
type IIA brane configuration presents a number of interesting phenomena, 
like explicit realization of chiral symmetries \cite{hanbro}, localization 
of chiral matter in the direction $x^6$ \cite{hzd8}, flavour doubling 
\cite{bhkl}, the effect of the type IIA cosmological constant, and 
transitions changing the number of flavours \cite{hzd8}.

The model has a simple T-dual description in terms of a set of IIB D3-branes 
sitting at an orbifold singularity, in the presence of a suitable set of 
D7-branes. We will show how the phenomena observed in the IIA side are 
reproduced in the IIB picture. Surprisingly enough, most of these exotic 
effects have a standard realization in the IIB setup.

\medskip

\begin{figure}
\centering
\epsfxsize=3.5in
\hspace*{0in}\vspace*{.2in}
\epsffile{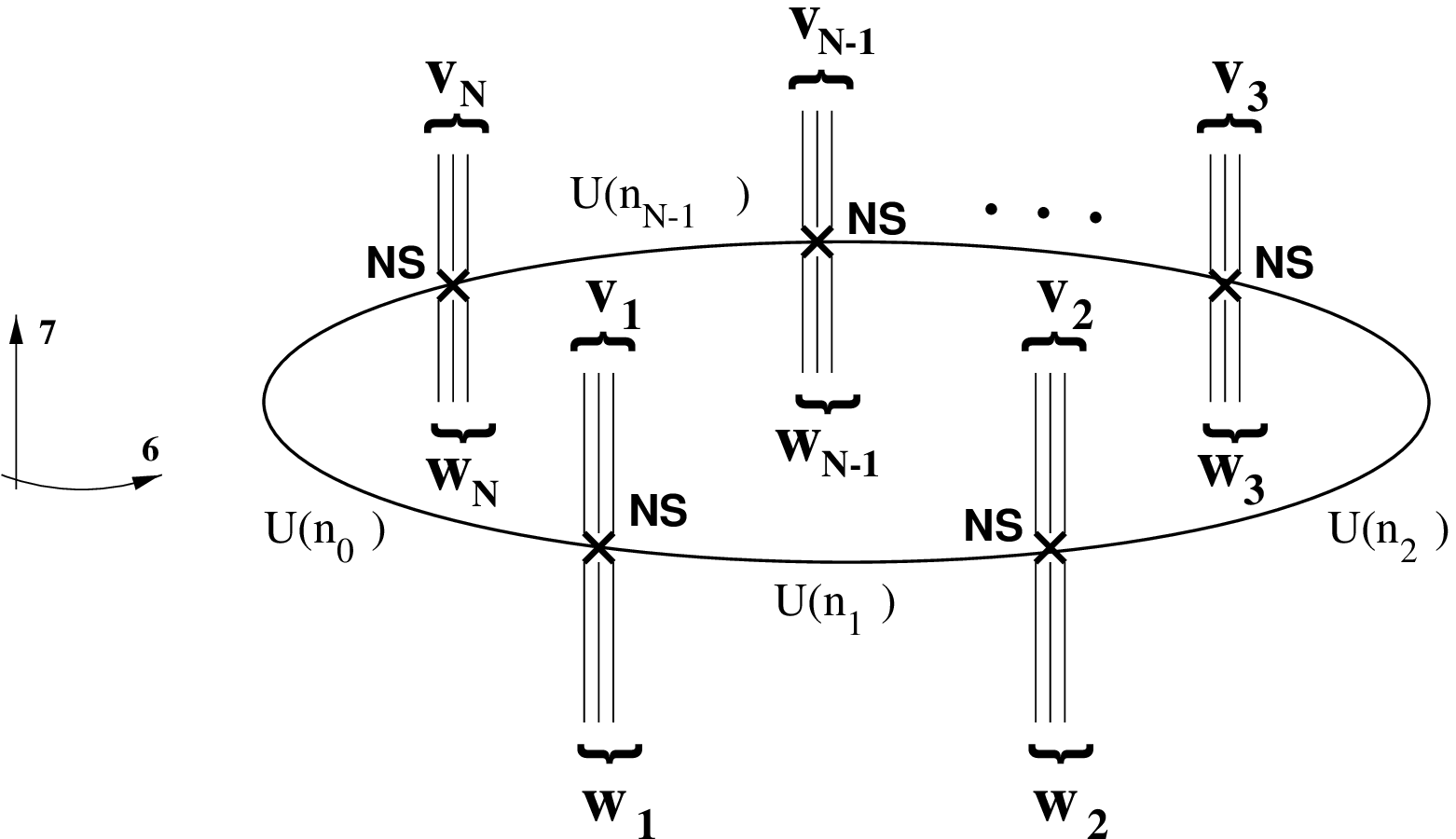}
\caption{\small Type IIA brane configuration corresponding to an $\NN=2$ 
elliptic model with a set of D6$'$-branes introducing $\NN=1$ chiral 
matter. In this picture, the crosses represent NS-branes and the thin 
vertical lines represent half D6$'$-branes. The D4-branes stretch between 
the NS-branes along the direction 6. For the sake of clarity, the D4-branes 
have been drawn at the origin in the Coulomb branch. This 
configuration presents the phenomenon of `flavour doubling': each half 
D6$'$-branes provides two chiral fundamental flavours, one for each gauge 
factor bounded by the corresponding NS brane.
} 
\label{fig:doubl0} 
\end{figure}

We start with a elliptic model \cite{wit4d} with $N$ NS branes (along the 
directions 012345) and $n_i$ D4 branes (along 01236) in the $i^{th}$ interval, 
suspended between NS-branes. The low-energy field theory on the D4-branes 
has the following gauge group and $\NN=1$ matter content
{\small
\beqa
 & \prod_{i=1}^N SU(n_i)\times U(1)  \nonumber\\
Q_{i,i+1} & \quad (\fund_i,\antifund_{i+1}) \nonumber\\
{\tilde Q}_{i+1,i} & \quad(\antifund_i,\fund_{i+1}) \\
\Phi_i & \quad {\rm Adj}_{\,i} \nonumber
\label{specellip}
\eeqa
}
and there is a $\NN=2$ superpotential
\beqa
W=\sum_{i=1}^N \; [\; \Phi_i Q_{i,i+1}{\tilde Q}_{i+1,i} - 
\Phi_{i+1}{\tilde 
Q}_{i+1,i} Q_{i,i+1} \; ]
\label{supellip}
\eeqa
Now we add a set of D6$'$ branes stretching along 0123457, such that they 
have semi-infinite extent along 7 and end on the NS-branes. We place 
$v_i$ `upper' ({\em i.e.} along positive $x^7$) and $w_i$ `lower' 
(negative $x^7)$ such half-D6$'$ on top of the $i^{th}$ NS-brane. The 
configuration and our conventions are shown in figure \ref{fig:doubl0}.

The presence of the D6$'$-branes breaks the supersymmetry of the 
configuration to four supercharges, $\NN=1$ in the four-dimensional field 
theory on the D4-branes. Since the $i^{th}$ such half-brane is sitting 
exactly at the boundary between the $(i-1)^{th}$ and the $i^{th}$ gauge 
factors, each half-D6$'$-brane gives a chiral flavour to each factor. 
This is the `flavour doubling' phenomenon \cite{bhkl}. So we have the 
additional $\NN=1$ matter
{\small
\beqa
\begin{array}{ccccc}
T^{w}_{i} \; : & \quad  (n_i,{\ov {w}}_i) \quad & ; & 
\quad  {\tilde T}^{w}_{i-1} \; : & \quad ({\ov {n}}_{i-1},w_i) \\ 
T^{v}_{i-1} \; : & \quad (n_{i-1},{\ov {v}}_i) \quad & ; &
\quad {\tilde T}^{v}_{i}\; : & \quad ({\ov {n}}_i,v_i) 
\end{array}
\label{spectes}
\eeqa
}
where the second entries specify the representation with respect to the gauge 
group $\prod_{i=1}^N SU(v_i)$$\times \prod_{i=1}^N SU(w_i)$ on the half 
D6$'$-branes, interpreted as global chiral symmetries from the field theory 
viewpoint \cite{hanbro}. Following \cite{bhkl}, there is also a $\NN=1$ 
superpotential
\beqa
\sum_{i=1}^N \; [\; Q_{i-1,i} T^{w}_{i}{\tilde T}^w_{i-1} - {\tilde Q}_{i,i-1} 
{\tilde T}^{v}_{i-1} T^{v}_{i} \; ]
\label{superptes}
\eeqa
which ensures that, when the NS brane is removed (that is, when the 
bifundamentals get a vev), two of the four kinds of  chiral multiplets 
become massive and we are left with the familiar situation  of each 
D6$'$-brane giving one vector-like flavour to only one gauge group. 
Notice that for D6$'$-branes there is no coupling between the flavours and 
the adjoint multiplets $\Phi_i$. Such coupling would not be consistent with 
the chiral symmetries mentioned above.

The configuration also contains states arising from strings stretched 
between upper and lower half D6$'$-branes on the same NS-brane. These are
six-dimensional fields, which transform in bifundamental representations 
$(v_i,{\ov w}_i)$. They parametrize the possibility of recombining upper 
and lower half D6$'$-branes and moving them off the NS-brane. 

Cancellation of non-abelian gauge anomalies imposes constraints on $w_i$, 
$v_i$. The anomaly for the $SU(n_i)$ gauge factor is proportional to  
$w_i-w_{i+1}-v_i+v_{i+1}$, so the anomalies vanish when 
\beqa
w_i-v_i=C_0
\label{anomorbi}
\eeqa
with $C_0$ is a constant independent of $i$. Notice that the matter content 
is non-chiral \footnote{Moreover, it can be organized in $\NN=2$ multiplets. 
However, the superpotential interactions in (\ref{superptes}) preserve 
only $\NN=1$ supersymmetry.}.

In the type IIA configuration anomaly cancellation is a consequence of the 
conservation of RR charge. Equation (\ref{anomorbi}) states that for $C_0=0$ 
the number of upper and lower half-D6$'$-branes must be equal. This is the 
usual condition of RR charge conservation \cite{bk1}. When $C_0\neq 0$ there 
is an $i$-independent mismatch between the number of upper and lower 
half-D6$'$-branes. This situation is consistent when the type IIA background 
has a non-zero cosmological constant \cite{hzd8} proportional to the mismatch 
$C_0$. Notice that even though the resulting field theories are 
non-chiral, the configurations have in general localized chiral matter.

A related interesting phenomenon was studied in \cite{hzd8}, where the 
introduction of D8-branes (along 012345689) in the configuration allowed 
to change the value of the cosmological constant. A change from $C_0$ to 
$C_0+1$ is achieved by crossing a D8-brane from a very large positive value 
of $x^7$ to a very large negative value. Consistency with the equation 
(\ref{anomorbi}) above requires either a decrease in the number of upper 
half D6$'$-branes, or an increase in the number of lower ones. This is 
achieved by the well-known brane creation effect \cite{hw}, which in this 
case implies that, at each NS-brane, either one upper half D6$'$-brane 
disappears (if it was attached at the crossing D8-brane) or one lower half 
D6$'$-brane is created. From the viewpoint of the four-dimensional field 
theory, the net effect is a sudden change in the number of flavours for each 
gauge group.

\medskip

In the remainder of this section we focus on a T-dual type IIB realization 
of these field theories, as a set of D3-branes probing an orbifold 
singularity, in the presence of D7-branes. We show how the different 
phenomena arising from the type IIA brane dynamics receive a simple 
interpretation in the type IIB picture.

Let us T-dualize the IIA model along the direction $x^6$, so that the 
NS-branes transform into a $N$-centered Taub-NUT space \cite{oovafa}, and the 
D4-branes become a set of D3 brane probes. We first consider the model 
without D6$'$-branes. A quite detailed account of properties of this 
T-duality can be found in \cite{kls}. As usual, at the origin of the baryonic 
and mesonic branches the centers of the Taub-NUT coincide and give a 
$\IC^2/\IZ_N$ singularity, at which D3-branes sit. Therefore in the following 
we center on the local structure, or equivalently take the limit in which the 
asymptotic Taub-NUT radius goes to infinity, and work with D3-branes at the 
orbifold singularity. The space is $\IC^2/\IZ_N \times \IC$, with 
the generator $\theta$ acting as
\beqa
\theta \; :\; (z_1,z_2,z_3) \rightarrow (e^{2\pi 
i\frac{1}{N}} z_1, e^{-2\pi i\frac{1}{N}} z_2, z_3)
\label{actionzn}
\eeqa
However, we will keep in mind the decompactification limit we have taken, 
since it may introduce subtleties to be discussed below.

The construction of the field theory in the type IIB setup is rather 
familiar in the absence of the fundamental flavours \cite{dm}. We start 
with a set of D3-branes in flat space $\IC^3$ and mod out by the geometric 
action (\ref{actionzn}), embedded on the Chan-Paton factors of D3-branes 
through the matrix
\beqa
\gamma_{\theta,3}=\diag (1_{n_0},e^{2\pi i\frac{1}{N}}1_{n_1}, \ldots, 
e^{2\pi i\frac{N-1}{N}}1_{n_{N-1}})
\label{cpthree}
\eeqa
The projection onto invariant fields, given by
{\small
\beqa
\begin{array}{ccccccc}
V=\gamma_{\theta,3} V \gamma_{\theta,3}^{-1} & \rightarrow & 
\prod_i U(n_i) & 
; & Z_1=e^{2\pi i/N}\gamma_{\theta,3} Z_1 \gamma_{\theta,3}^{-1} & 
\rightarrow & \sum_{i=1}^N (\fund_i,\antifund_{i+1}) \\
Z_3=\gamma_{\theta,3} Z_3 \gamma_{\theta,3}^{-1} & 
\rightarrow & {\rm Adj}_{\; i} & ; &
Z_2=e^{-2\pi i/N}\gamma_{\theta,3} Z_2 \gamma_{\theta,3}^{-1} & 
\rightarrow & (\antifund_i,\fund_{i+1}) \\
\end{array}
\eeqa
}
reproduces the spectrum of the field theory (\ref{specellip}) 
\footnote{The $U(1)$ factors appearing in the orbifold theory (except for 
the decoupled diagonal combination) are made massive by the couplings 
$B\wedge \tr F$ on the D3-brane world-volume. This phenomenon is the 
T-dual of the freezing of $U(1)$'s observed in the IIA brane configuration 
\cite{wit4d}.}. The interactions (\ref{supellip}) are recovered from the 
orbifold interaction, inherited from the $\NN=4$ superpotential $W=Z_3 
[Z_1,Z_2]$. Notice that the $\NN=1$ multiplets $V$, $Z_3$ and $Z_1$, $Z_2$ 
form $\NN=2$ vector and hypermultiplets respectively.

The introduction of flavours is more interesting, and has not appeared in 
the literature. It can be accomplished  by the introduction of D7-branes 
in the configuration. We will denote by D7$_i$ the D7-branes transverse 
to the $i^{th}$ complex plane. It is easy to see that D7$_3$ branes 
preserve the $\NN=2$ supersymmetry and thus they correspond to D6-branes 
in the IIA configuration. On the other hand, D7$_1$- and  D7$_2$-branes 
break the supersymmetry to $\NN=1$ and seem the appropriate objects to 
match the IIA D6$'$-branes. We now show that the  spectrum (\ref{spectes}) 
is reproduced by the string states in the 3-7$_i$, 7$_i$-3 sectors if we 
make the following choice of Chan-Paton matrices
\beqa
\gamma_{\theta,7_1} & = & \diag (e^{\pi i \frac{1}{N}} 1_{w_1},e^{\pi i 
\frac{3}{N}} 1_{w_2},\ldots, e^{\pi i\frac{2N-1}{N}}1_{w_N})
\label{cp7}
\eeqa
and analogously for the D7$_2$-branes \footnote{The unusual property that 
$\Tr \gamma_{\theta^N,3}=1$ while $\Tr \gamma_{\theta^N,7_1}=\Tr 
\gamma_{\theta^N,7_2}=-1$ is actually required for consistency of the model 
(let us point out that if D7$_3$-branes are included, they must have $\Tr 
\gamma_{\theta^N,7_3}=1$). This follows for instance from the analysis in 
\cite{bl} of a system of D9- and D5$_i$-branes (we refer to D5-branes 
wrapping the complex plane $z_i$ as D5$_i$-branes). The simultaneous rotation 
by $2\pi$ in the complex planes $z_1$, $z_2$ was shown to act with opposite 
signs on the Chan-Paton factors of D9-, D5$_3$-branes vs. D5$_1$-, 
D5$_2$-branes. By T-dualizing along $z_1$, we recover our system of D3- and 
D7-branes, with $\theta^N$ acting with opposite signs on D3-, D7$_3$ vs, 
D7$_1$-, D7$_2$-branes.}, replacing the numbers of entries $w_i$ by $v_i$. 

The spectrum in the 3-7$_1$, 3-7$_2$ sectors is obtained through the 
projections 
{\small
\beqa
3-7_1: & \quad \lambda  = e^{-\frac{\pi i}{N}} \; \gamma_{\theta^k,3}\lambda 
\gamma_{\theta^k,7_1}^{-1}  & \rightarrow (n_i,{\ov w}_i)
\nonumber \\
7_1-3: & \quad \lambda  = e^{-\frac{\pi i}{N}}\; \gamma_{\theta^k,7_1}\lambda 
\gamma_{\theta^k,3}^{-1} & \rightarrow ({\ov n}_{i-1},w_i)
\nonumber \\
3-7_2: & \quad \lambda  = e^{\frac{\pi i}{N}} \; \gamma_{\theta^k,3}\lambda 
\gamma_{\theta^k,7_2}^{-1} & \rightarrow (n_{i-1},{\ov v}_{i})
\nonumber \\
7_2-3: & \quad \lambda  = e^{\frac{\pi i}{N}}\; \gamma_{\theta^k,7_2}\lambda 
\gamma_{\theta^k,3}^{-1} & \rightarrow ({\ov n}_i,v_i)
\label{proj37}
\eeqa
}
Notice that each D7$_i$-brane provides one chiral flavour to two gauge 
groups of the four-dimensional field theory. This statement is the T-dual 
of the flavour doubling effect in the IIA configuration, and appears 
automatically in the IIB setup. The superpotential (\ref{superptes}) is 
also reproduced in the orbifold model, by the couplings 
(3-3)$_i$(3-7$_i$)(7$_i$-3) among 3-3 states associated to $Z_i$ and 
states in the 3-7$_i$ and 7$_i$-3 sectors. Also, no coupling to the 
adjoint multiplets appears.

This proposal also reproduces several other features of the IIA configuration. 
For instance, the global chiral symmetries in the four-dimensional field 
theory correspond to the gauge symmetries in the D7-brane world-volumes. 
This suggests the proposal that the half-D6$'$-branes in the IIA side map 
to complete D7-branes in the IIB side. In fact, this observation has been 
already made in \cite{kg} in a related context.

The seemingly striking feature that the IIA half D6$'$-branes, which 
originally span the same directions, end up as D7-branes spanning different 
complex planes in $\IC^2/\IZ_N$ deserves some comment. This point is better 
understood if we consider the ALE space to be slightly blown-up, and is 
described by $xy=\prod_{i=1}^N (v-v_i)$, where $v_i$ correspond to the 
positions of the T-dual NS-branes in 89. In this case, the D6-branes are not 
split in halves, and map to D7-branes located at $v=0$, so that they wrap the 
complex curve $xy=\prod_{i=1}^N v_i$. When the NS-branes are brought back to 
$v=0$, this curve degenerates to $xy=0$, which represents two complex planes 
intersecting at one point. The D7-branes wrapping each of these correspond to 
different half D6$'$-branes.

As further evidence for this identification, we study the string consistency 
conditions in the type IIB configuration, namely cancellation of tadpoles for 
twisted RR fields. More concretely, consistency only requires the cancellation 
of D7-brane tadpoles, since those corresponding to the D3-branes yield no 
inconsistency, since the RR flux can escape to infinity along the complex 
plane $z_3$. The D7-brane tadpoles appear from cylinder diagrams, and give 
the constraint
\beqa
\Tr \gamma_{\theta^k,7_1}-\Tr \gamma_{\theta^k,7_2}=0 \quad {\rm for}\; 
k\neq 0
\label{tadorb}
\eeqa
These conditions ensure the cancellation of non-abelian anomalies in the 
four-dimensional field theory, as can be checked by expressing (\ref{tadorb}) 
in terms of $v_i$ and $w_i$ (see e.g. \cite{abiu} for extensive application 
of this trick). Notice the general solution to the constraints is given by 
\vspace*{-.5cm}
\beqa
v_i-w_i=C_0
\label{czero}
\eeqa
where the constant $C_0$ is independent of $i$, and counts the difference 
between the number of D7$_1$- and D7$_2$ branes. Comparing with 
(\ref{anomorbi}), we obtain the result that this number in the IIB 
configuration is mapped to the cosmological constant on the IIA side.

Finally, let us comment on a subtle point concerning deformations of the 
sets of D7-branes. One can check that  the states in the 7$_1$-7$_2$ sector 
in the IIB orbifold correspond to the bifundamentals $(v_i,{\ov w}_i)$ 
living on the D6$'$-brane world-volume in the IIA configuration. The 
corresponding deformations are identical to those in \cite{kg}: In the 
type IIA side, the bifundamental parametrizes the possibility of 
recombining upper and lower half-D6$'$-branes and moving them off the NS 
brane; in the IIB side, it parametrizes the deformation of two transverse 
D7-branes into a single smooth one \cite{sen}.

On the other hand, the 7$_1$-7$_1$ and 7$_2$-7$_2$ sector give rise to a 
set of bifundamental multiplets $(w_i,{\bar w}_{i+1})$, $(v_i,{\bar 
v}_{i+1})$, etc, which have no IIA counterpart. The deformations 
associated to these IIB states correspond to combining different 
fractional D7-branes into a dynamical one and moving it away from the 
orbifold singularity. These deformations seem to be absent in the type 
IIA description \footnote{In a particular case, ref. \cite{kg} proposed a 
deformation of two, say upper, half-D6$'$-branes into a parabola. However 
this type of deformation would does not solve the problem in the general 
case, since the IIB deformation would require the participation of half 
D6$'$-branes located at different NS-branes.}.

Notice that there is a crucial difference between the 7$_1$-7$_2$ and 
7$_i$-7$_i$ states. While the former are localized at a point in the orbifold 
space, the latter spread out on complex planes. This fact implies that 
the 7$_i$-7$_i$ states are sensitive to the asymptotic radius of the 
Taub-NUT space which is the true T-dual of the IIA brane configuration. 
More concretely, these 7$_i$-branes wrap the compact direction of the 
Taub-NUT space, hence the `asymptotic' Wilson lines (actually self-dual 
connections asymptotically flat) may give masses to these states, 
inversely proportional to the Taub-NUT radius (so they become massless in 
the ALE limit). These states should therefore be mapped to  massive states 
in the corresponding type IIA configuration, with masses proportional to 
the IIA $x^6$ radius. In fact, the IIA model contains reasonable 
candidates for such states, as open strings stretched between half 
D6$'$-branes sitting at different NS-branes. It would be interesting to 
obtain further support for this picture. We would like to stress that 
this interpretation implies that  the removal of a dynamical D7-brane from 
the orbifold singularity is {\em not} a supersymmetric deformation of 
the type IIB configuration for finite asymptotic radius of the Taub-NUT 
space.

To finish this section, we would like to comment on the T-dual version of 
the transitions changing the number of flavours in the four-dimensional 
field theory. As explained above, in the IIA picture this process takes 
place when D8-branes along 012345689 are moved in $x^7$ and cross the 
configuration. So it will be useful to identify the T-dual of the D8-brane 
in the IIB picture. On general grounds, the T-dual of a D8-brane should 
correspond to a D7-brane wrapped on a holomorphic curve $\Sigma$ in the 
Taub-NUT space, so as to preserve the same supersymmetries as the remaining 
branes. Let us represent the Taub-NUT as slightly blown-up for the sake of 
clarity, $xy=\prod_{i=1}^N (v-v_i)$, where $v$ corresponds to the coordinates 
89. The curve $\Sigma$ is parametrized by $v$, and it is located at a 
specific value of $x/y$. The modulus of $x/y$ is related to the $x^7$ 
position of the D8-brane, whereas its phase corresponds to the coordinate 
T-dual to $x^6$. Recall that the D7-branes T-dual to the D6$'$-branes 
wrap the curve $\Delta$ given by $v=0$. Notice that $\Delta$ and 
$\Sigma$ are transverse and intersect once, as is required to reproduce 
the spectrum of 6$'$-8 states ({\em i.e.} the number of DN+ND directions 
is preserved).

Now let us argue how the transition in which flavours of the field theory 
are created/anihilated occurs. When the curve $\Sigma$ is moved in the 
direction $x^7$ in the Taub-NUT, eventually it will hit its centers (notice 
that even though $\Sigma$ is localized in the $S^1$ coordinate, the collision 
with the centers cannot be avoided, since the compact direction shrinks to 
zero size at the centers). Clearly it is enough to consider the crossing of 
one center. During this crossing the non-trivial metric of the Taub-NUT near 
the center induces a change in the vortex number for the gauge field 
propagating on $\Sigma$. This change must be compensated by the creation of a 
brane that stretches between the Taub-NUT center and the D7-brane. More 
concretely, what has been created is a D7-brane tube stretching from the 
Taub-NUT center to the D7-brane, to which it joins smoothly \footnote{This 
process is geometrically identical to the M-theory lift of the crossing of a 
type IIA NS-brane with a D6-branes, in which a D4-brane is created. Moreover, 
if appropriate directions are compactified on circles, the processes are 
related by T-duality.}. When the D7-brane T-dual to the D8-brane is carried 
away to infinity, one is left with an infinite D7-brane, localized in $v$ at 
the position of the corresponding Taub-NUT center. This is precisely the type 
of curve wrapped by the T-dual of one upper or lower D6$'$-brane, that 
is, it become a D7$_1$ or D7$_2$ (depending on whether the original D8-brane 
moved upwards or downwards in $x^7$) in the orbifold limit. The creation 
of this D7-brane explains the net change in the number of flavours in the 
field theory. It also implies that the constant $C_0$ above has changed 
in one unit, even though we have not found a more direct derivation of 
this fact. 

Since the orientifold models studied in this paper are obtained by a 
projection of the orbifold states into $\IZ_2$-invariant fields, all the 
comments in this section apply to the orientifold models to come. Therefore, 
in principle we study the models without additional chiral flavours (unless 
they are required for consistency), and briefly discuss more general 
configurations in Section~5.

\section{Introduction of O6$'$-planes in the type IIA configurations}

There is a very simple way to obtain field theories with $\NN=1$ 
supersymmetry using type IIA brane configurations. We start with a 
configuration of NS-branes (along 012345) and D4-branes (along 01236) leading 
to $\NN=2$ field theories \cite{wit4d}, and introduce orientifold planes that 
break half of the supersymmetries. This can be done introducing orientifold 
six-planes along 0123457 (referred to as O$6'$-planes), fixed under 
$(-1)^{F_L}I_{689}$ (where $I_{689}$ is a reflection in the coordinates 689). 
These brane configurations present a number of interesting properties related 
to chiral symmetries and chiral matter content in the four-dimensional field 
theory. Such are the splitting of D6$'$-branes in halves, or the change of 
sign in the orientifold RR charge as it crosses a NS-brane. This type of 
configurations has been studied in the non-compact $x^6$ case in a 
number of papers (for instance see \cite{lll,egkt,bhkl}), which provide 
the rules to read off the spectrum and interactions of the  resulting $\NN=1$ 
field theories. We will be interested in models where the direction $x^6$ has 
been compactified on a circle, therefore the configuration contains two 
orientifold planes at opposite sides on the circle.

In the following we make a classification of the possible models depending 
on the number $N$ of NS branes \footnote{We will usually work with $\IZ_2$ 
invariant configurations on the $\IZ_2$ covering space of the orientifold. 
Thus $N$ denotes the number of NS-branes in this covering space.} and their 
positions with respect to the O$6'$-planes. The classification is analogous 
to that in \cite{pu} (see also \cite{uranga,ehn}) for $\NN=2$ theories. 
Notice 
that we do not include D6$'$-branes in the models below, unless they are 
required for consistency. More general configurations of D6$'$-branes are 
discussed in Section~5.

\subsection{Odd number of NS branes, $N=2P+1$}

In these configurations, we have $P$ pairs of two NS branes related by 
the $\IZ_2$ orientifold symmetry, and one unpaired NS brane which intersects
one of the O$6'$-planes. Notice that the NS brane actually divides the 
O$6'$-plane in two halves, so the orientifold plane changes sign when it 
crosses the NS brane \cite{ejs}. Due to conservation of RR charge in this 
crossing, we have to add 8 half D6$'$-branes on the negatively charged side 
of the O$6'$-plane. This is the `fork' configuration appeared in 
\cite{bhkl,lll,egkt} (see also \cite{park,lll2}), which yields a chiral 
matter content. 

\begin{figure}
\centering
\epsfxsize=5in
\hspace*{0in}\vspace*{.2in}
\epsffile{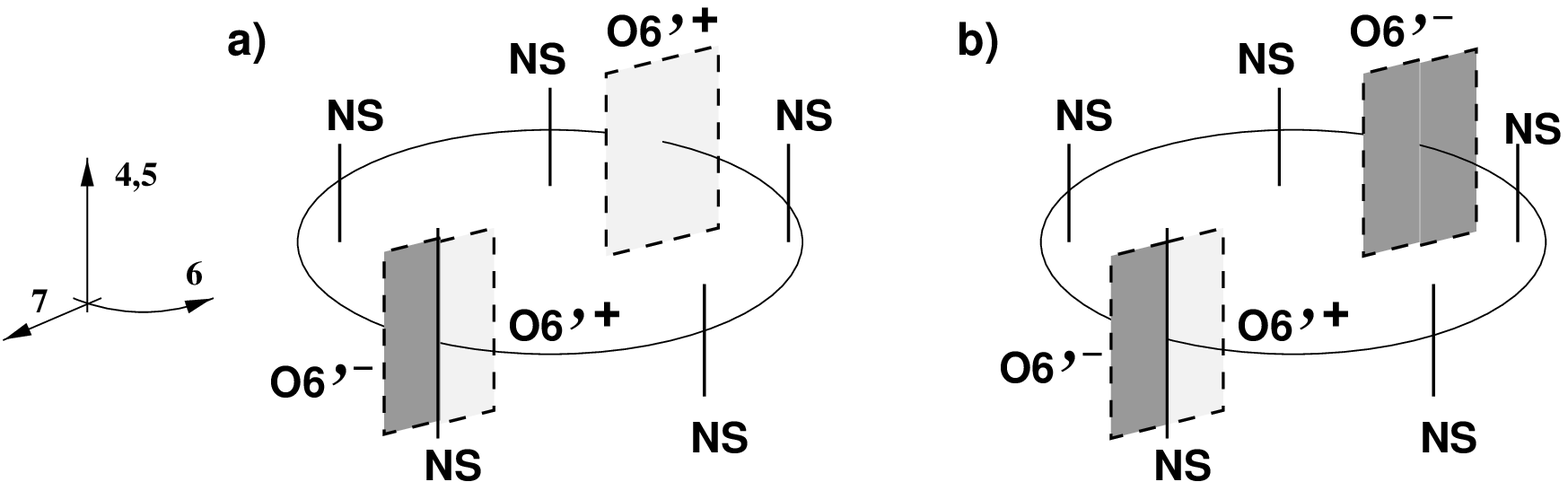}
\caption{\small Typical $\NN=1$ brane configurations obtained by introducing 
O$6'$-planes in $\NN=2$ elliptic models with odd number of NS-branes. 
Notice the fork configuration, where the O$6'$-plane changes sign when it 
crosses the NS-brane. The change of sign is depicted by a change of color. 
The eight half D6$'$-branes stuck on the negatively charged half of the 
orientifold are not shown in the picture.} 
\label{fig:oddn1} 
\end{figure}

The charge of the remaining O$6'$-plane can be chosen of either sign, so 
for a given $N$ there are two possible configurations, which are depicted
in figure~\ref{fig:oddn1}. To 
give an example of the field theories that are obtained from these brane 
configurations, let us consider the fork-O$6\,'\,^{+}$ configuration, 
figure~\ref{fig:oddn1}a. The field theory gauge group and $\NN=1$ matter 
content is
{\small
\beqa
\begin{array}{ccc}
& SO(n_0)\times \prod_{i=1}^P U(n_i) & \\
Q_{i,i+1} & (\fund_i,\antifund_{i+1}) & i=0,\ldots,P-1 \\
{\tilde Q}_{i+1,i} & (\fund_{i+1},\antifund_{i}) & i=0,\ldots,P-1\\
S_0 & \Ysymm_0 & \\
\Phi_i & {\rm Adj}_{\; i} & i=1,\ldots,P \\
{\tilde S}_P & {\ov {\Ysymm}}_P & \\
A_P & \Yasymm_P & \\
T_{P}^a & \fund_{P} & a=1,\ldots,8
\end{array}
\label{specodd}
\eeqa
}
The superpotential is given by
\beqa
W & = & S_0 Q_{0,1} {\tilde Q_{1,0}} + \sum_{i=1}^{P-1}\, [\, - \Phi_i 
{\tilde Q}_{i,i-1} Q_{i-1,i} \; + \; \Phi_i Q_{i,i+1}{\tilde 
Q}_{i+1,i}\, ]\, + \nonumber\\ 
 & & - \Phi_P {\tilde Q}_{P,P-1} Q_{P-1,P} + {\tilde S}_P T_P T_P
\label{supodd}
\eeqa
The field theory for the fork-O$6\,'\,^{-}$ configuration, 
figure~\ref{fig:oddn1}b, is obtained by replacing $SO(n_0)$ by $USp(n_0)$, 
and the symmetric representation $S_0$ by an antisymmetric $A_0$.

\medskip

\subsection{Even number of NS branes, $N=2P$}

When the number of NS-branes is even, $N=2P$, there are two possible 
$\IZ_2$-invariant arrangements of NS branes, depending on whether or 
not there are NS-branes coinciding with the O$6'$-planes. 

\medskip

\subsubsection{Models without NS branes on top of the orientifold planes}

\begin{figure}
\centering
\epsfxsize=4in
\hspace*{0in}\vspace*{.2in}
\epsffile{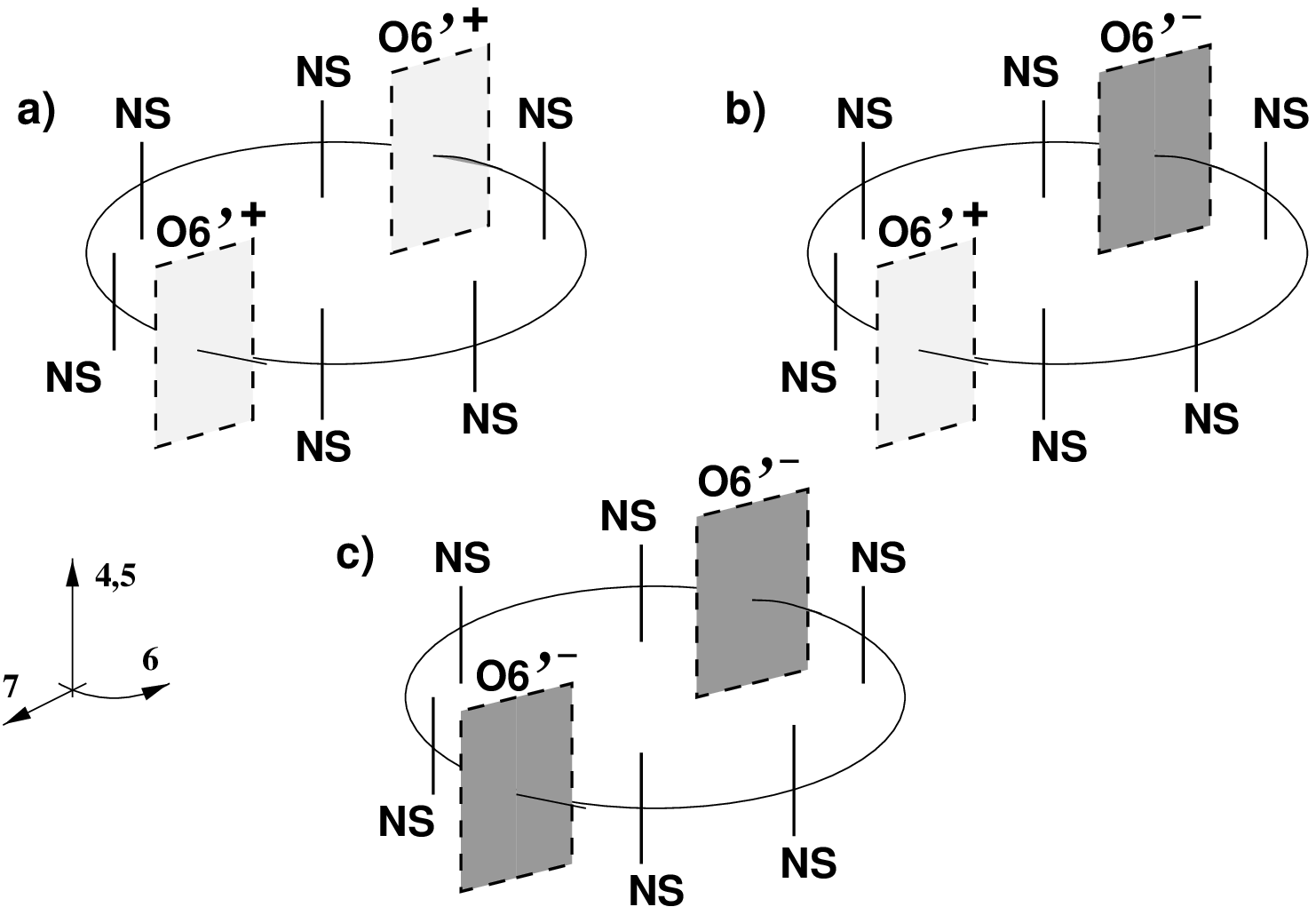}
\caption{\small The $\NN=1$ brane configurations with O$6'$-planes and an 
even number of NS-branes, in the case in which no NS-brane coincides with 
the orientifold planes.} 
\label{fig:evn1vs} 
\end{figure}

The configurations contain $P$ pairs of NS-branes related by the $\IZ_2$ 
symmetry. Since no NS-branes intersects the O6$'$-planes, there are no 
fork configurations and the theories will be non-chiral. As shown in figure 
\ref{fig:evn1vs}, there are three possible choices for the signs of the 
O$6'$-planes.

The field theories can be read off directly from the brane picture. For 
instance, for the \posp configuration in figure~\ref{fig:evn1vs}a, we have
{\small
\beqa
\begin{array}{ccc}
& SO(n_0)\times \prod_{i=1}^{P-1} U(n_i)\times SO(n_P) & \\
Q_{i,i+1} & (\fund_i,\antifund_{i+1}) & i=0,\ldots,P-1 \\
{\tilde Q}_{i+1,i} & (\fund_{i+1},\antifund_{i}) & i=0,\ldots,P-1\\
S_0 & \Ysymm_{\,0} & \\
\Phi_i & Adj_{\,i} & i=1,\ldots,P-1 \\
S_P & \Ysymm_P & 
\end{array}
\label{speceven1}
\eeqa
}
The superpotential is given by
\beqa
W  =  S_0 Q_{0,1} {\tilde Q_{1,0}} + \sum_{i=1}^{P-1}\, [\, - \Phi_i 
{\tilde Q}_{i,i-1} Q_{i-1,i} \; + \; \Phi_i Q_{i,i+1}{\tilde 
Q}_{i+1,i}\, ]\,  - S_P {\tilde Q}_{P,P-1} Q_{P-1,P}
\label{supeven1}
\eeqa
The field theory corresponding to the \posnegp configuration in 
figure~\ref{fig:evn1vs}b is obtained from the above one by changing {e.g.} 
$SO(n_0)$ by $USp(n_0)$ and the symmetric $S_0$ by an antisymmetric $A_0$. 
The field theory for the configuration in \negp in figure~\ref{fig:evn1vs}c 
is obtained by changing both orthogonal factors to symplectic ones, and 
both symmetric representations to antisymmetric ones.
 
\medskip

\subsubsection{Models with NS branes coinciding with the O$6'$-planes}

\begin{figure}
\centering
\epsfxsize=5in
\hspace*{0in}\vspace*{.2in}
\epsffile{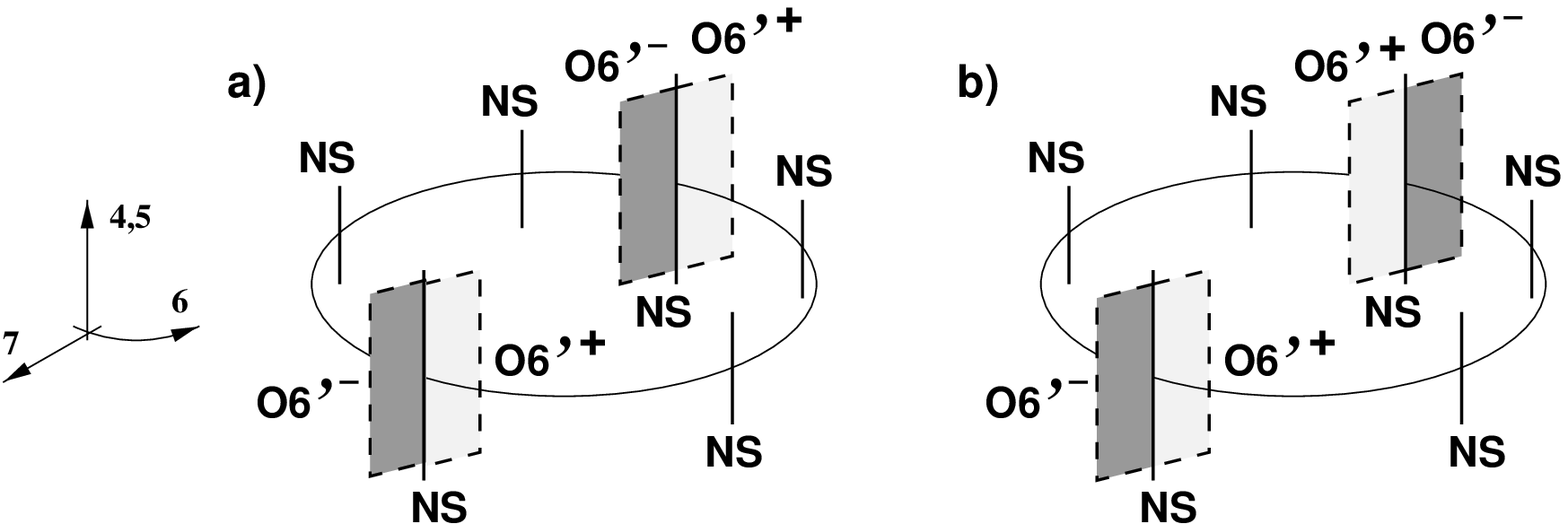}
\caption{\small The $\NN=1$ brane configurations with O$6'$-planes and an 
even number of NS branes, in the case with NS branes coinciding with 
the orientifold planes.} 
\label{fig:evn1nvs} 
\end{figure}

These configurations contain $P-1$ pairs of NS-branes, and two unpaired 
NS-branes. The latter are stuck at the two O6$'$-planes, which are therefore
split in halves and give rise to two fork configurations. There are 
two possible models for a given $N$, shown in figure~\ref{fig:evn1nvs}, 
which only  differ in the orientation of the forks in $x^7$. We will denote 
the two relative orientations as `antiparallel' and `parallel'. 

Considering for instance the case of `parallel' forks, the field theory 
obtained is given by
{\small
\beqa
\begin{array}{ccc}
& \prod_{i=1}^P U(n_i) & \\
Q_{i,i+1} & (\fund_i,\antifund_{i+1}) & i=1,\ldots,P-1 \\
{\tilde Q}_{i+1,i} & (\fund_{i+1},\antifund_{i}) & i=1,\ldots,P-1\\
S_1 & \Ysymm_1 & \\
{\tilde A}_1 & \bYasymm_{\, 1} & \\
{\tilde T}_{1,a} & \antifund_1 & a=1,\ldots,8 \\
\Phi_i & {\rm Adj}_{i} & i=1,\ldots,P \\
{\tilde S}_P & {\ov {\Ysymm}}_P & \\
A_P & \Yasymm_P & \\
T_{P,a} & \fund_P & a=1,\ldots,8 \\
\end{array}
\label{specfork1}
\eeqa
}
The superpotential is given by
\beqa
W  = & S_1 {\tilde T}_1 {\tilde T}_1 + \Phi_1 Q_{1,2} {\tilde Q}_{2,1} + 
\sum_{i=2}^{P-1} \; [\, - \Phi_i {\tilde Q}_{i,i-1} Q_{i-1,i} \; 
+ \; \Phi_i Q_{i,i+1}{\tilde Q}_{i+1,i}\, ]\, + \nonumber \\
& - \Phi_P {\tilde Q}_{P,P-1} Q_{P-1,P} + {\tilde S}_P T_P T_P
\label{supfork1}
\eeqa
Changing the orientation of a fork simply conjugates the corresponding 
representations $S$, ${\tilde A}$ and ${\tilde T}$. Even though the 
difference between the two configurations may seem rather subtle, the models 
are different and should be treated as such. For instance, notice the 
particular case of having only two NS branes, where one of the 
configurations is chiral and the other is non-chiral (even though there 
is chiral matter  localized at points in the circle).

\medskip

We have completed the classification of type IIA configurations obtained 
by adding O6$'$-planes to the $\NN=2$ elliptic models. In the following 
subsection we construct the same field theories in a T-dual IIB 
singularity picture.

\section{The type IIB orientifold construction}

In this section we provide a type IIB T-dual construction of the field 
theories studied in the previous section, in terms of D3-branes probing 
certain orientifolds of $\IC^2/\IZ_N$. We will show that the 
classification of the different orientifold projections reproduces the 
classification of type IIA models. This T-duality map gives a type IIB 
singularity realization of the different phenomena observed in the IIA 
brane construction, giving rise to chiral symmetries and chiral matter.
These features of the field theory arise in a natural fashion in the 
orientifold construction.

\medskip

Our goal in this section is to provide a construction of these field theories 
in the type IIB side. Since in the type IIA setup the models were obtained 
by introducing orientifold planes in a $\NN=2$ elliptic model, we expect 
the type IIB realization to correspond to an orientifold projection on 
the system of D3-branes at an orbifold singularity $\IC^2/\IZ_N \times 
\IC$, with the generator $\theta$ of $\IZ_N$ acting as
\beqa
(z_1,z_2,z_3) \rightarrow (e^{2\pi i/N} z_1,e^{-2\pi i/N}z_2,z_3)
\label{orbifold}
\eeqa
A difference with the $\NN=2$ orientifold models studied in  \cite{pu}
is that in our case the orientifold projection must break half of the 
supersymmetries, yielding $\NN=1$ field theories on the D3-brane 
world-volume. Thus, a suitable orientation reversing element is 
$\Omega_1\equiv\Omega (-1)^{F_L}R_1$ (rather than the $\NN=2$ preserving 
$\Omega_3\equiv\Omega(-1)^{F_L}R_3$). Here $R_i$ changes the sign of the 
complex coordinate $z_i$, leaving the others invariant.

Our proposal is that the field theories will be realized in the 
world-volume of a system of D3-branes at the orientifold singularities 
obtained by modding the orbifold (\ref{orbifold}) by the world-sheet 
orientation reversing action $\Omega_1$. This is actually suggested by 
T-dualizing the IIA configurations along $x^6$ (for a detailed 
description of this T-duality in a particular case, see \cite{kg}). 
Namely, the NS-branes along 012345 transform into the orbifold above (if
we identify the 45, 67, 89 planes with, say, the $z_3$, $z_2$, $z_1$ 
complex planes), the D4-branes turn into D3-brane probes, and the type IIA 
orientifold action $\Omega(-1)^{F_L}I_{689}$ transforms into $\Omega 
(-1)^{F_L}R_1$.

\medskip

In the following section we turn to the classification and explicit 
construction of these type IIB orientifolds. We show that the field theories 
on the D3-branes actually reproduce the field theories in the previous 
section. Moreover, we discuss how the orientifold structure on the IIA 
side is neatly encoded in the structure of the IIB orientifold group. This 
strongly supports the T-duality argument sketched above, and also 
provides a singularity realization of the fork configurations yielding 
chiral matter. We also mention that the consistency conditions of the IIB 
orientifolds (cancellation of twisted tadpoles) imply the cancellation of 
gauge anomalies in the D3-brane world-volume theory (the corresponding 
tadpole computations are however postponed until Section~5). These models 
provide an interesting realization  of how an inconsistency in a 
background is detected as an anomaly on a D-brane probe.

\subsection{Odd order case, $N=2P+1$}

\subsubsection{Construction}

Let us consider the case where the orientifold group is ${\IZ_N} + 
\Omega_1 {\IZ_N}$, {\em i.e.} the orientifold projection has the structure
\beq
(1+\theta+\theta^2+\cdots +\theta^{N-1})(1+\Omega_1)
\label{orodd}
\eeq
where $\Omega_1\equiv\Omega(-1)^{F_L}R_1$

We will consider the following choice of Chan-Paton matrices for the 
D3-branes.
\beqa
&\gamma_{\theta,3}\, =\, {\rm diag} (1_{n_0},e^{2\pi i\frac{1}{N}} 
1_{n_1}, \ldots, e^{2\pi i\frac{P}{N}} 1_{n_P}, e^{2\pi 
i\frac{P+1}{N}}1_{n_P}, \ldots, e^{2\pi i\frac{2P}{N}} 1_{n_1})  \\
\nonumber \\
&{\footnotesize \gamma_{\Omega_1,3} = \pmatrix{
1_{n_0} &  &  &  &  &  & \cr
&  &  &  &  &  & 1_{n_1} \cr
&  &  &  &  & \cdots &  \cr
&  &  &  & 1_{n_P} &  &  \cr
&  &  & 1_{n_P} &  &  &  \cr
&  & \cdots &  &  &  &  \cr
& 1_{n_1} &  &  &  &  &  \cr
} \; {\rm or} \;
\gamma_{\Omega_1,3} =\pmatrix{
\varepsilon_{n_0} &  &  &  &  &  &  \cr
  &  &  &  &  &  & 1_{n_1}  \cr
  &  &  &  &  & \cdots &  \cr
  &  &  &  & 1_{n_P} &  &  \cr
  &  &  & -1_{n_P} &  &  &  \cr
  &  & \cdots &  &  &  &  \cr
  & -1_{n_1} &  &  &  &  &  \cr
}} \nonumber
\eeqa
where $\varepsilon_{n_0}=\pmatrix{ 0 & 1_{n_0/2} \cr -1_{n_0/2} & 0}$.
These matrices satisfy the group law and all algebraic consistency conditions 
imposed by the structure of the orientifold group. The two possibilities 
for $\gamma_{\Omega_1,3}$, symmetric or antisymmetric, correspond to 
choosing the $SO$ or $Sp$ orientifold projection on the D3-branes.

It is a simple exercise to compute the 3-3 spectrum that arises from the 
projections imposed by the above matrices. For instance, for the $SO$ 
projection, we obtain the following gauge group and matter content (we 
have already taken into account the disappearance of the $U(1)$ factors)
{\small
\beqa
& SO(n_0) \times SU(n_1) \times SU(n_P) & \nonumber \\
& \sum_{i=0}^{P-1} \; [\; (\fund_i,\antifund_{i+1}) + 
(\antifund_i,\fund_{i+1}) \;]\; + \Yasymm_P + {\ov {\Ysymm}}_P 
+ \Ysymm_0 + \sum_{i=1}^{P} {\rm Adj}_{\, i} &
\eeqa
}
It reproduces the spectrum (\ref{specodd}) of the field theory corresponding 
to the IIA configuration figure~\ref{fig:oddn1}a, studied in section~3. 
This was expected, since the T-duality argument implies the order of the 
orbifold group matches the number of NS-branes in the IIA picture. The 
superpotential interactions in the IIB orientifold also reproduce the IIA 
result (\ref{supodd}).

Notice that the spectrum above is actually missing some of the states in 
the IIA brane configuration, namely the eight fundamental flavours arising 
from the half D6$'$-branes in the fork. The solution to this issue comes 
from the fact that the IIB orientifold above is not consistent yet, since 
it contains non-cancelled tadpoles (see section 5.1). This inconsistency 
is manifest in the field theory, since the $SU(n_P)$ factor is anomalous. 
The tadpoles can be 
cancelled by introducing D7 branes. The minimal choice to cancel the tadpoles 
(see eq. (\ref{tadp1})) is to place eight D7$_2$-branes with Chan-Paton 
matrices 
\beqa
\gamma_{\theta,{7_2}}=-{\bf 1}_{8} \quad ; \quad \gamma_{\Omega_1,7_2} = 
{\bf 1}_{8}
\eeqa
The 3-7$_2$ spectrum yields the 8 fundamentals of $U(n_P)$ required to cancel 
the anomalies and complete the matching with the spectrum obtained from the 
type IIA brane configuration. The interactions between 3-7$_2$ states and 3-3 
states (more concretely, those associated to the second complex plane) in the 
orientifold provides the coupling between the fundamentals and symmetric of 
$SU(n_P)$. Also, the $SO(8)$ symmetry on the D7$_2$-brane world-volume agrees 
with the $SO(8)$ symmetry of the half D6$'$-branes in the IIA picture. The 
fact that whole D7$_2$-branes are T-dual to half D6$'$-branes confirms our 
expectations from the analysis in section~2. Further evidence is provided in 
section 5.1, where we study models with more complicated choices of D7-brane 
Chan-Paton matrices, and their interpretation in terms of type IIA picture.

Another interesting feature is that the geometry of the type IIB 
orientifold encodes the change of sign of the IIA O6$'$-plane. In 
figure \ref{fig:oddn1}a, for negative values of $x^7$ both O6$'$-planes are 
positively charged, while for positive values of $x^7$ their charges are 
opposite. This translates into different behaviours of the IIB orientifold 
projection with respect to the two complex planes in $\IC^2/\IZ_N$. In fact, 
the orientifold group (\ref{orodd}) contains the element 
$\Omega(-1)^{F_L}R_1$, and the model contains an O7$^+$-plane 
wrapped along $z_2$. This is the T-dual to the two positively charged 
O6$'$-planes in the positive $x^7$ region in the IIA construction. On the 
other hand, the IIB orientifold group does not contain $\Omega(-1)^{F_L} 
R_2$, so there is no RR charge along the complex plane$z_1$. This 
corresponds to the fact that the O6$'$-plane charges compensate in the 
negative $x^7$ region in the IIA picture.

The orientifold obtained with the antisymmetric $\gamma_{\Omega_1,3}$ can be 
analyzed analogously. It provides a construction of the field theory 
corresponding to the IIA configuration in figure \ref{fig:oddn1}b. 

\medskip

As a final comment, let us mention that there is another consistent 
structure for the orientifold group, given by
\beq
(1+\theta+\theta^2+\cdots +\theta^{N-1})(1+\alpha\Omega_1),
\eeq
with $\alpha: (z_1,z_2,z_3)\to (e^{\pi i/N}z_1,e^{-\pi i/N}z_2,z_3)$. 
However, it does not provide new models. In order to see that, recall that 
$\theta^P\alpha=R_1R_2$, so the projection above is equivalent to a 
projection using $\Omega_2\equiv\Omega(-1)^{F_L}R_2$ (without $\alpha$), 
which is equivalent to the one above.

\medskip

\subsubsection{Comparison with $\NN=2$ models}

The $\NN=1$ orientifolds just constructed differ from the $\NN=2$ models 
of \cite{pu} only in the use of $\Omega_1$ instead of $\Omega_3$ as the 
orientifold action. This apparently innocent change has nevertheless a 
dramatic effect in the field theory. In particular, it is responsible for 
the appearance of the fork structure, and consequently, of chiral matter.
In what follows we compare both projections, and discuss how the $\Omega_1$ 
orientifold yields the fork matter content. The projections are given in the 
following table 
{\small
\renewcommand{\arraystretch}{1.25}
\begin{center}
\begin{tabular}{|c|c|c|}
\hline
& {\bf $\NN=2$ projection} & {\bf $\NN=1$ projection} \\  
\hline
{\bf Vector mult:} & 
$\quad V=-\gamma_{\Omega_3} V^{T}\gamma_{\Omega_3}^{-1}\quad$ & 
$\quad V=-\gamma_{\Omega_1} V^{T}\gamma_{\Omega_1}^{-1}\quad$ 
\\
{\bf Ch.mult. $Z_1$:} & 
$\quad Z_1=+\gamma_{\Omega_3} Z_1^{T} \gamma_{\Omega_3}^{-1} \quad$ & 
$\quad Z_1=-\gamma_{\Omega_1} Z_1^{T} \gamma_{\Omega_1}^{-1} \quad$ 
\\
{\bf Ch.mult. $Z_2$:} & 
$\quad Z_2=+\gamma_{\Omega_3} Z_2^{T} \gamma_{\Omega_3}^{-1} \quad$ & 
$\quad Z_2=+\gamma_{\Omega_1} Z_2^{T} \gamma_{\Omega_1}^{-1} \quad$ 
\\
{\bf Ch.mult. $Z_3$:} & 
$\quad Z_3=-\gamma_{\Omega_3} Z_3^{T} \gamma_{\Omega_3}^{-1} \quad$ & 
$\quad Z_3=+\gamma_{\Omega_1} Z_3^{T} \gamma_{\Omega_1}^{-1} \quad$\\
\hline
\end{tabular}
\end{center}
}
Recall that before the orientifold projection, the $\NN=1$ multiplets $V$ 
and $Z_3$ pair up to form $\NN=2$ vector multiplets. Analogously, the chiral 
multiplets $Z_1$, $Z_2$ form $\NN=2$ hypermultiplets. This $\NN=2$ 
structure is preserved by the projection $\Omega_3$, while the projection 
$\Omega_1$ introduces relative signs that break the structure of $\NN=2$ 
multiplets. The effects of these signs are most manifest in two sectors of 
the field theory which we now analyze:

Consider first the two $\NN=1$ multiplets $V$, $Z_3$ in the $\NN=2$ vector 
multiplet in the adjoint of $U(n_0)$ in the orbifold theory, before the 
orientifold projection. From the above equations (for a specific choice 
of  $\gamma_{\Omega_i}$, say, symmetric), we see $\Omega_3$ projects both 
multiplets to adjoints of $SO(n_0)$, in agreement with $\NN=2$ supersymmetry. 
On the other hand, $\Omega_1$ projects the $\NN=1$ vector multiplet $V$ to 
the adjoint of $SO(n_0)$, and the chiral multiplet $Z_3$ to the symmetric 
representation. This nicely parallels the results in the IIA brane 
configuration \cite{bhkl}, where $n_0$ D4-branes stretched between two 
NS-branes, in the presence of an O6$^+$-plane give rise to a $SO(n_0)$ gauge 
group with one antisymmetric chiral multiplet \cite{landlop}, and in the 
presence of an O6$'\,^+$-plane give a $SO(n_0)$ gauge group and one 
symmetric chiral multiplet.

Now we consider the projection at the other end of the chain of gauge 
factors,  and look at what kind of two-index representations of $U(n_P)$ 
are obtained. For the projections $\Omega_3$, we obtain one chiral 
multiplet in the symmetric and one in the conjugate symmetric representation, 
in agreement with $\NN=2$ supersymmetry. For the projection $\Omega_1$ we 
obtain one chiral multiplet in the antisymmetric and one in the conjugate 
symmetric representation. This is basically the fork configuration that 
appears in the T dual IIA configuration. 

As mentioned above, the spectrum in the fork configuration also includes 
chiral flavours coming from eight half D6$'$-branes, required to conserve the 
RR charge. The $\NN=2$ IIA configuration, however, does not require the 
presence of D6-branes for consistency. These features are also reproduced in 
the IIB orientifold construction, as is revealed by studying the dependence 
of the Klein bottle tadpoles with the volume $V_3$ of the complex plane $z_3$, 
which we momentarily imagine to be compact. The projection $\Omega_3$ allows 
winding states in the direction $z_3$, which give (after Poisson resumming)
a tadpole inversely proportional to $V_3$. The tadpole vanishes in the 
non-compact limit, and the model is consistent without the introduction of 
D7-branes. The projection $\Omega_1$, however, projects out winding states 
and allows for momentum states. These generate a tadpole proportional to $V_3$, 
which does not vanish in the non-compact limit, and must be cancelled by a 
suitable set of D7-branes. These provide the fundamental flavours required to 
complete the fork spectrum in the D3-brane field theory.

\medskip

\subsection{Even order case, $N=2P$}

When the order of the orbifold action is even, the group contains a $\IZ_2$ 
twist, $\theta^P$. The orientifold action on the closed string exchanges 
oppositely twisted sectors \cite{polchinski}, and so maps the $\IZ_2$ 
twisted sector to itself. There is an arbitrary choice of sign in this 
map, which determines the symmetry of the NS-NS and R-R states that 
survive the orientifold projection. By open-closed duality, the choice of 
sign imposes a constraint on the Chan-Paton matrices which act on the open 
strings, and determines the vector structure of the gauge bundle on the 
corresponding D-branes (for further details see \cite{blpssw, polchinski, 
bi1}). For instance, for D7$_1$-branes we have
\beqa
\gamma_{\theta^P,7_1}=\mp \gamma_{\Omega_1,7_1} \gamma_{\theta^P,7_1}^T 
\gamma_{\Omega_1,7_1}^{-1} 
\eeqa
with the upper (lower) sign when the $\IZ_2$ RR twisted states surviving the 
orientifold projection are antisymmetric (symmetric) combinations of left 
and right movers \footnote{This relation follows from the result in 
\cite{polchinski} for D9-branes by T-duality along $z_1$.}. These two cases 
correspond to $\Tr \gamma_{\theta^N,3}=\pm 1$, respectively, so we will use 
this fact to refer to both kinds of models.

\subsubsection{Models with $\Tr \gamma_{\theta^N,3}=+1$} 

There are two different structures for the orientifold group. Consider 
first the orientifold projection
\beq
(1+\theta+\theta^2+\cdots +\theta^{N-1})(1+\Omega_1)
\label{gorientone}
\eeq
The Chan-Paton matrices for the D3 branes are
\beqa
\gamma_{\theta,3} & = & \diag(1_{n_0}, e^{2\pi i\frac{1}{N}} 1_{n_1}, 
\cdots,e^{2\pi i \frac{P-1}{N}} 1_{n_{P-1}}, e^{2\pi i\frac{P}{N}} 1_{n_P}, 
e^{2\pi i \frac{P+1}{N}} 1_{n_{P-1}}, \ldots, e^{2\pi i\frac{2P-1}{N}} 
1_{n_1}) \nonumber\\
\gamma_{\Omega_1,3} & = & {\footnotesize \left( \begin{array}{cccccccc}
1_{n_0} & & & & & & &  \\
  & & & & & & & 1_{n_1} \\
  & & & & & &\cdots &  \\
  & & & & & 1_{n_{P-1}}& &  \\
  & & & & 1_{n_P} & & & \\
  & & & 1_{n_{P-1}}& & & & \\
  & & \cdots & & & & & \\
  & 1_{n_1} & & & & & &  \end{array}  \right)}
\label{soproj} 
\eeqa
These matrices satisfy all algebraic consistency conditions. The 
corresponding 3-3 spectrum is \vspace*{-.5cm}
{\small
\beqa
& SO(n_0) \times SU(n_1) \times \ldots \times SU(n_{P-1})\times SO(n_P) &
\nonumber \\
& \sum_{i=0}^{P-1} \; [\; (\fund_i,\antifund_{i+1}) + 
(\antifund_i,\fund_{i+1}) \;]\; + \Ysymm_0 + \sum_{i=1}^{P-1} 
{\rm Adj}_{\, i} + \Ysymm_P &
\eeqa
}
This reproduces the spectrum of the field theory (\ref{speceven1}) 
corresponding to the IIA configuration \ref{fig:evn1vs}a, studied in 
Section~3. The superpotential interactions (\ref{supeven1}) are also 
reproduced in the orientifold. The field theory corresponding to the \negp 
IIA brane configuration, figure \ref{fig:evn1vs}c, is reproduced by an 
analogous IIB orientifold using an antisymmetric version for 
$\gamma_{\Omega_1}$. 

In these models, the consistency of the IIA brane configuration without 
D6$'$-branes suggests that the type IIB orientifold must be consistent 
without the addition of D7-branes. In fact, this is confirmed by a 
computation of the Klein bottle tadpoles, which are found to vanish (see 
section 5.2.1).

Notice that the orientifold group (\ref{gorientone}) contains both the 
elements $\Omega (-1)^{F_L}R_1$ and $\Omega (-1)^{F_L}R_2$. Therefore, 
there is RR charge along the complex planes parametrized by $z_1$ 
and $z_2$. Under the T-duality, this implies that the IIA configuration 
contains identically charged upper half O6$'$-planes, as well as 
identically charged lower half O6$'$-planes. This is indeed the case for 
our candidate T-dual IIA models. 

\medskip

There is another possibility for the orientifold group 
\beq
(1+\theta+\theta^2+\cdots +\theta^{N-1})(1+\alpha\Omega_1)
\label{gorientitwo}
\eeq
with $\alpha:(z_1,z_2,z_3) \to (e^{\pi i/N} z_1, e^{-\pi i/N} z_2, z_3)$.
As we show below, this structure provides new models. Let us consider the 
following D3-brane Chan-Paton matrices
{\small
\beqa
\gamma_{\theta,3} & = & {\rm diag} (1_{n_0},e^{2\pi i\frac{1}{N}} 1_{n_1},
\ldots, e^{2\pi i\frac{P-1}{N}} 1_{n_{P-1}}, e^{2\pi i \frac{P}{N}} 1_{n_P},
e^{2\pi i\frac{P+1}{N}} 1_{n_{P-1}},\ldots,e^{2\pi i\frac{2P-1}{N}} 1_{n_1}) 
\nonumber\\ 
\gamma_{\alpha\Omega_1,3} & = & {\scriptsize \pmatrix{
1_{n_0}  &  &  &  &  &  &  &  \cr
  &  &  &  &  &  &  & \alpha 1_{n_1} \cr
  &  &  &  &  &  & \cdots &  \cr
  &  &  &  &  & \alpha^{P-1}1_{n_{P-1}} &  &  \cr
  &  &  &  & \alpha^P \varepsilon_{n_P}  &  &  &  \cr
  &  &  & \alpha^{-(P-1)} 1_{n_{P-1}} &  &  &  &  \cr
  &  & \cdots &  &  &  &  &  \cr
  & \alpha^{-1} 1_{n_1} &  &  &  &  &  &  \cr
}}
\eeqa
}
where $\alpha=e^{2\pi i\over {2N}}$. These matrices satisfy 
the group law and all algebraic consistency conditions. The spectrum of 
massless 3-3 states is
{\small
\beqa
& SO(n_0) \times SU(n_1) \times \ldots \times SU(n_{P-1})\times USp(n_P) &
\nonumber \\
& \sum_{i=0}^{P-1} \; [\; (\fund_i,\antifund_{i+1}) + 
(\antifund_i,\fund_{i+1}) \;]\;+ \Ysymm_0 + \sum_{i=1}^{P-1} {\rm Adj}_{\, i} + \Yasymm_P &
\eeqa
}
and reproduces the complete field theory spectrum and interactions of the 
IIA configuration in figure \ref{fig:evn1vs}b. A nice feature of the type IIB 
orientifold we have constructed is that the matrix $\gamma_{\alpha\Omega_1}$ 
is rather unique, neither symmetric nor antisymmetric. The fact that this 
orientifold group structure provides only one class of models agrees with the 
type IIA construction, where the \posnegp and \negposp configurations are 
equivalent. To complete the discussion, let us mention that all dangerous 
tadpoles vanish in this model (see section 5.2.1), so it is consistent not 
to include D7-branes. 

Finally, let us mention that the orientifold group (\ref{gorientitwo}) 
contains neither $\Omega(-1)^{F_L}R_1$ nor $\Omega(-1)^{F_L}R_2$, and 
therefore there is no RR charge along the complex planes in $\IC^2/\IZ_N$.
This suggests that the O6$'$-plane charges in the T-dual type IIA 
configuration will cancel both in the positive and in the negative $x^7$ 
regions, as indeed happens for the model in figure \ref{fig:evn1vs}b.

\medskip

\subsubsection{Models with $\Tr \gamma_{\theta^N,3}=-1$} 

As in the theories in the previous subsection, there are two possible 
choices for the orientifold group. Let us start by considering
\beq
(1+\theta+\theta^2+\cdots +\theta^{N-1})(1+\Omega_1)
\label{gorone}
\eeq
We take the following Chan-Paton matrices for the D3 branes
{\small
\beqa
& \gamma_{\theta,3}  =  \diag(e^{\pi i\frac{1}{N}}1_{n_1},
e^{\pi i\frac{3}{N}}1_{n_2}, \cdots, e^{\pi i\frac{2P-1}{N}}1_{n_P},
e^{\pi i\frac{2P+1}{N}}1_{n_P}, \cdots,e^{\pi i\frac{4P-3}{N}}1_{n_2},
e^{\pi i\frac{4P-1}{N}}1_{n_1}) \quad \quad \\
\nonumber\\
& \gamma_{\Omega_1,3}  =  {\scriptsize \left( \begin{array}{cccccccc}
  & & & & & & & 1_{n_1} \\
  & & & & & &1_{n_2} &  \\
  & & & & & \cdots& &  \\
  & & & & 1_{n_P} & & & \\
  & & & 1_{n_P}& & & & \\
  & & \cdots & & & & & \\
  &  1_{n_2}& & & & & & \\
  1_{n_1} & & & & & & &  \end{array}  \right)}
\eeqa
}
(In this case using an antisymmetric version of $\gamma_{\Omega_1}$ does 
{\em not} produce new models). The matrices satisfy all algebraic 
consistency conditions. The 3-3 spectrum is
{\small
\beqa
& SU(n_1)\times \ldots \times SU(n_P) & \nonumber\\
& \bYasymm_1 + \Ysymm_1 + \sum_{i=1}^{P-1} \;[\; (\fund_{i},\antifund_{i+1})
+ (\antifund_i,\fund_{i+1}) \;]\; + \Yasymm_P + {\ov {\Ysymm}}_P + 
\sum_{i=1}^P {\rm Adj}_{\, i} &
\label{parall}
\eeqa
}
This reproduces most of the spectrum of the field theory (\ref{specfork1}), 
which corresponds to the IIA configuration in figure \ref{fig:evn1nvs}, with 
`parallel' forks. As discussed in section 5.2.2, this orientifold contains 
non-vanishing Klein bottle tadpoles. The minimal choice to cancel them 
(see eq.(\ref{tadnovecone})) is to introduce a set of D7$_2$-branes with 
Chan-Paton matrices
\beqa
\gamma_{\theta,7_2} = \diag (1_8,-1_8) \quad ; \quad 
\gamma_{\Omega_1,7_2}={\bf 1}_{16}
\eeqa
The 3-7 spectrum provides the eight antifundamental flavours for $SU(n_1)$ 
and eight fundamental flavours for $SU(n_P)$ required to cancel the 
anomalies and complete the matching with the spectrum (\ref{specfork1}).
The superpotential interactions in the IIB orientifold neatly reproduce 
the IIA result (\ref{supfork1}). 

We can also show how the orientifold group (\ref{gorone}) encodes the 
structure of O6$'$-planes in the T-dual IIA model. The presence of 
$\Omega(-1)^{F_L}R_1$ and $\Omega(-1)^{F_L}R_2$ in the orientifold group 
implies that there is RR charge along both complex planes in 
$\IC^2/\IZ_N$. This implies the IIA model has net O6$'$-plane charge both 
in the upper and lower regions in $x^7$, as indeed is the case when the 
forks have the same orientation (parallel). Further agreement comes from 
checking that the IIB model contains one O7$^+$ and one O7$^-$. This can
be seen by directly computing the RR charge or looking at the orientifold 
actions on D7$_1$- and D7$_2$-branes (see subsection 5.2.2).

\medskip

There is a second possible structure for the orientifold projection, namely
\beq
(1+\theta+\theta^2+\cdots +\theta^{N-1})(1+\alpha\Omega_1)
\label{gortwo}
\eeq
The D3-brane Chan-Paton matrices have the general structure
\beqa   
\gamma_{\theta,3} & = & {\rm diag} (e^{\pi i\frac{1}{N}}1_{n_1},
e^{\pi i\frac{3}{N}}1_{n_2},\ldots, e^{\pi i\frac{2P-1}{N}}1_{n_{P}},
e^{\pi i\frac{2P+1}{N}}1_{n_P},\ldots, e^{\pi i\frac{4P-3}{N}} 1_{n_2},
e^{\pi i\frac{4P-1}{N}} 1_{n_1}) \nonumber \\
\gamma_{\alpha\Omega_1,3} & = & {\footnotesize \pmatrix{
  &  &  &  &  & e^{\pi i\frac{1}{2N}} 1_{n_1}   \cr
  &  &  &  & \cdots &    \cr
  &  &  & e^{\pi i\frac{2P-1}{2N}} 1_{n_P}&  &    \cr
  &  & e^{\pi i\frac{2P+1}{2N}} 1_{n_P} &  &  &    \cr
  & \cdots &  &  &  &    \cr
e^{\pi i\frac{4P-1}{2N}} 1_{n_1}  &  &  &  &  &    \cr
}}
\eeqa
and satisfy all the requirements for algebraic consistency. The 3-3 spectrum 
obtained from the relevant projection is analogous to (\ref{parall}), 
differing only in the conjugation of the representations $\bYasymm_1$, 
$\Ysymm_1$ to $\Yasymm_1$, ${\ov {\Ysymm}}_1$. This realizes the field theory 
corresponding to the IIA configuration in figure \ref{fig:evn1nvs}, with 
`antiparallel' forks. As in the previous model, the missing states arise in 
the 3-7 sector when one introduces the D7-branes required to cancel the 
tadpoles in the orientifold. In this case, the minimal choice of D7$_1$, 
D7$_2$ Chan-Paton factors is (see eq. (\ref{tadnovectwo}))
\beqa
& \gamma_{\theta,7_1}={\bf 1}_8 \quad &  \gamma_{\alpha\Omega_1,7_1}={\bf 
1}_8 \nonumber \\
& \gamma_{\theta,7_2}=-{\bf 1}_8 \quad &  \gamma_{\alpha\Omega_1,7_2}={\bf 
1}_8
\eeqa
This provides eight chiral fundamental flavours for $SU(n_1)$ and $SU(n_P)$, 
precisely the amount required to reproduce the type IIA spectrum and cancel 
the field theory gauge anomalies.

In this case, the group (\ref{gortwo}) implies there is no RR charge along 
the complex planes in $\IC^2/\IZ_N$. Therefore, we expect no net RR charge 
along either positive or negative $x^7$ in the T-dual IIA configuration. 
This is indeed the case for antiparallel forks.

This concludes our classification and construction of $\NN=1$ orientifolds 
of $\IC^2/\IZ_N$ models. The results are summarized in table~\ref{table1}, 
which shows how the different orientifolds we have constructed reproduce 
on the D3-brane probes the field theories we had classified in the type 
IIA setup.
{\small
\begin{table}[htb]
\renewcommand{\arraystretch}{1.25}
\begin{center}
\begin{tabular}{|c||c||c|}
\hline
$\NN=1$ & Type IIB orientifold & Type IIA configuration \\
\hline\hline
Odd $N$ & ${\IZ_N}+\Omega_1{\IZ_N}^*$ & Fig. \ref{fig:oddn1} : 
O6$'^{\,+}$-NS$_1$-...-NS$_P$-fork \\
\cline{3-3}
$N=2P+1$ & & \hspace{.5cm} O6$'^{\,-}$-NS$_1$-...-NS$_P$-fork \\
\hline\hline
 & ${\IZ_N}+\Omega_1{\IZ_N}^*$ & Fig. \ref{fig:evn1vs}a : 
O6$'^{\,+}$-NS$_1$-...-NS$_P$-O6$'^{\,+}$ \\
\cline{3-3}
 & $\Tr \gamma_{\theta^N,3}=+1$ & Fig. \ref{fig:evn1vs}c : 
O6$'^{\,-}$-NS$_1$-...-NS$_P$-O6$'^{\,-}$ \\
\cline{2-3} 
 & ${\IZ_N}+\alpha\Omega_1{\IZ_N}$ & Fig. \ref{fig:evn1vs}b : 
O6$'^{\,-}$-NS$_1$-...-NS$_P$-O6$'^{\,+}$ \\
Even $N$  & $\Tr \gamma_{\theta^N,3}=+1$ & \\
\cline{2-3}
$N=2P$ & ${\IZ_N}+\Omega_1{\IZ_N}$ & Fig. \ref{fig:evn1nvs} : 
fork-NS$_1$-...-NS$_{P-1}$-fork \\
 & $\Tr \gamma_{\theta^N,3}=-1$ & (parallel forks) \\
\cline{2-3}
 & ${\IZ_N}+\alpha\Omega_1{\IZ_N}$ & Fig. \ref{fig:evn1nvs} :
fork-NS$_1$-...-NS$_{P-1}$-fork \\
 & $\Tr \gamma_{\theta^N,3}=-1$ & (antiparallel forks) \\
\hline 
\end{tabular}
\end{center}
\caption{\small This table shows the different type IIB $\NN=1$ orientifolds 
of $\IC^2/\IZ_N$, constructed in Section~4, along with their T-dual type 
IIA brane configurations, classified in Section~3. An asterisk in the 
orientifold group indicates there are two possible projections ($SO$ or $Sp$) 
on the D3-branes.}
\label{table1}
\end{table}
}

\section{More general D7-brane structure} 

Here we would like to consider the orientifold models of Section~4 with a 
more general choice of Chan-Paton matrices for the D7-branes, and discuss 
the computation of the tadpole conditions. We will also describe the 
corresponding type IIA brane configurations, which are those in Section~3 
with additional D6$'$-branes. These models will provide the type IIB 
realization of several phenomena in the IIA construction. The results in this 
discussion parallel those in the orbifold models of section~2. 

\subsection{Odd order case, $N=2P+1$}

These are the models constructed in section 4.1. For concreteness, we will 
center on the case of $SO$ projection on the D3-branes. The models with $Sp$ 
projection can be studied analogously. An interesting configuration is 
obtained by using D7$_1$- and D7$_2$-branes (with $z_1$ and $z_2$ as their 
transverse directions, respectively), with Chan-Paton matrices 
{\small
\beqa
\gamma_{\theta,7_1} & = & \diag(e^{\pi i \frac{1}{N}} 1_{w_1}, \ldots ,
e^{\pi i \frac{2P-1}{N}} 1_{w_{P}}, e^{\pi i \frac{2P+1}{N}} 1_{w_{P+1}}, 
e^{\pi i \frac{2P+3}{N}} 1_{w_P},\ldots, e^{\pi i \frac{4P+1}{N}} 
1_{w_1})\quad 
\label{D7cpodd}
\eeqa
}
and an analogous $\gamma_{\theta,7_2}$, with the numbers of entries given 
by $v_i$ instead of $w_i$. We also take
\beqa
\gamma_{\Omega_1,7_1}  = {\scriptsize \left( \begin{array}{ccccccc}
  & & & & & & 1_{w_1} \\
  & & & & &\cdots &  \\
  & & & & 1_{w_P} & & \\
  & & & \varepsilon_{w_{P+1}}& & & \\
  & & -1_{w_P}& & & & \\
  & \cdots & & & & &  \\
  -1_{w_1}& & & & & & \\
\end{array}  \right)} \quad
\gamma_{\Omega_1,7_2}  = {\scriptsize \left( \begin{array}{ccccccc}
  & & & & & & 1_{v_1} \\
  & & & & &\cdots &  \\
  & & & & 1_{v_P} & & \\
  & & & 1_{v_{P+1}}& & & \\
  & & 1_{v_P}& & & & \\
  & \cdots & & & & &  \\
  1_{v_1}& & & & & & \\
\end{array}  \right) } \nonumber
\label{D7orodd}
\eeqa 
The symmetry of the matrices $\gamma_{\Omega_1,7}$ is not arbitrary: The 
orientifold requires the projection onto D7$_1$ (vs D7$_2$) to be opposite 
(similar) to the projection on D3-branes \footnote{This follows for instance 
from arguments in \cite{bl} (generalizing \cite{gp}) stating that, in a model 
with D9-, D5$_1$ and D5$_2$-branes the orientifold action on D5$_i$-branes is 
opposite to that on D9-branes. Our claim above follows after T-dualizing 
along $z_1$.}. In our case, D3- and D7$_2$-branes have a $SO$ projection, and 
D7$_1$-branes have a $Sp$ projection. 

The spectrum in the 3-7$_1$, 3-7$_2$ sectors is obtained through the 
projections (\ref{proj37}). Recalling that $\Omega_1$ relates the 
3-7$_i$ sector to the 7$_i$-3 sector, the resulting spectrum is
{\small
\beq
\sum_{i=1}^P \; [\; (n_i,{\ov w_i}) + ({\ov n_{i-1}},w_i)\; ] + 
({\ov n_P},w_{P+1}) + \sum_{i=1}^P \; [\; ({\ov n_i},v_i) + (n_{i-1},{\ov 
v_i})\; ] + (n_P,v_{P+1}) 
\label{spec37odd}
\eeq
}
where the second entry gives the representations under the D7-brane groups
$\prod_{i=1}^P SU(w_i)\times USp(w_{P+1})$ and $\prod_{i=1}^P SU(v_i) \times 
SO(v_{P+1})$. The superpotential is
\beqa
W = & \sum_{i=0}^{P-1} \; [ \;(n_i,{\ov n}_{i+1}) (n_{i+1},{\ov w}_{i+1}) 
(w_{i+1},{\ov n}_i) \; ] \; + \Yasymm_P ({\ov n}_P,w_{P+1})^2 + \nonumber\\
- & \sum_{i=0}^{P-1} \; [ \;({\ov n}_i,n_{i+1}) ({\ov n}_{i+1},v_{i+1}) 
({\ov v}_{i+1},n_i) \; ] \; - \Ysymm_P (n_P,v_{P+1})^2 + 
\label{superp1}
\eeqa

The integers $v_i$, $w_i$ are constrained by cancellation of twisted 
tadpoles, whose computation we sketch in what follows. The tadpoles can be 
extracted from the literature on six-dimensional compact orientifolds 
\cite{sag2,bs,gp,gjdp}. In particular, the expression relevant to our 
discussion can be 
directly taken from \cite{gjdp}, by changing only the zero mode structure. 
The twisted tadpoles arising from the cylinder, M\"obius strip and Klein 
bottle amplitudes are given by
\beqa
\cc & = & \sum_{k=1}^{N-1} (\Tr \gamma_{\theta^k,7_1})^2 + \sum_{k=1}^{N-1} 
(\Tr\gamma_{\theta^k,7_2})^2 -2 \sum_{k=1}^{N-1} \Tr \gamma_{\theta^k,7_1} 
\Tr\gamma_{\theta^k,7_2} \nonumber \\
\cm & = & -16 \sum_{k=1}^{N-1} \Tr(\gamma_{\theta^k \Omega_1,7_1}^T 
\gamma_{\theta^k \Omega_1,7_1}^{-1})  -16 \sum_{k=1}^{N-1}
\Tr(\gamma_{\theta^k \Omega_1,7_2}^T \gamma_{\theta^k \Omega_1,7_2}^{-1}) 
\nonumber \\ 
\ck & = & \sum_{k=1}^{N-1} 64
\eeqa
The trigonometric functions that usually appear in the numerator 
of these contributions are exactly cancelled against the trigonometric 
functions in the denominator that appear from the counting of zero modes. 
This feature is also shared by the remaining orientifold models.

Using the properties 
\beqa
\Tr(\gamma_{\theta^k \Omega_1,7_i}^T \gamma_{\theta^k \Omega_1,7_i}^{-1}) 
= \pm\Tr\gamma_{\theta^{2k},7_i}
\eeqa
(with the negative (positive) sign for D7$_1$- (D7$_2$-) branes) and the 
relation $\sum_{k} \Tr \gamma_{\theta^{2k},7_i}=\sum_k (-1)^k \Tr 
\gamma_{\theta^k,7_i}$, the tadpoles factorize as follows
\beqa
\sum_{k=1}^{N-1} \; [\; \Tr \gamma_{\theta^k,7_1} - \Tr 
\gamma_{\theta^k,7_2} + 8 (-1)^k  \; ]\; =0
\eeqa
The cancellation condition is therefore
\beq
\Tr\gamma_{\theta^k,7_1} -\Tr\gamma_{\theta^k,7_2}+8(-1)^k=0\; , \quad 
{\rm for} \;\; k\neq 0
\label{tadp1}
\eeq
Notice that the Chan-Paton matrices for D3-branes are irrelevant, since 
the corresponding RR tadpole can escape to infinity along the complex 
plane $z_3$.

These equations correspond to the conditions of cancellation of non-abelian 
anomalies in the D3-brane probe world-volume. This can be shown for 
instance by finding the general solution to (\ref{tadp1}), which is
\beqa
v_i=w_i+C_0 + 8\; \delta_{i,P+1}
\label{anom1}
\eeqa
Here $C_0$ is a constant independent of $i$. This equation states that the 
number of fundamentals and antifundamental 
multiplets is equal for each gauge group, save for $SU(n_P)$, which has an 
excess of eight fundamental flavours to cancel the anomaly generated by the 
antisymmetric and conjugate symmetric multiplets in the 3-3 sector.

\begin{figure}
\centering
\epsfxsize=4in
\hspace*{0in}\vspace*{.2in}
\epsffile{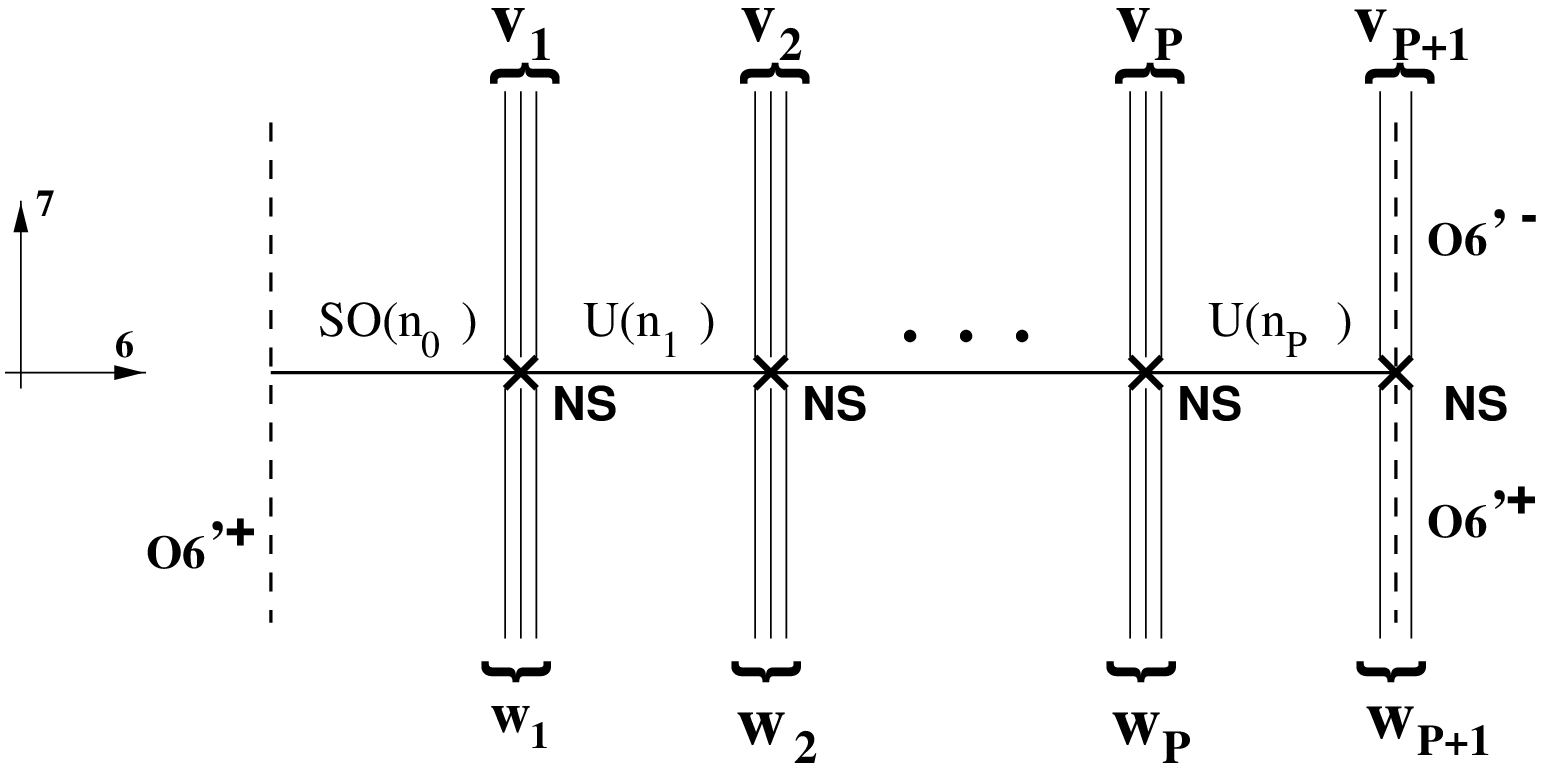}
\caption{\small The IIA brane configuration corresponding to the odd 
order IIB orientifold with a more general distribution of D7-branes. For 
clarity, in this picture we have depicted one copy of the interval and 
not its image by the orientifold $\IZ_2$ symmetry. NS branes are indicated 
by crosses, D4 branes by horizontal lines, the O$6'$-planes by vertical 
dashed lines, and the half-D$6'$-branes by vertical continuous lines, with 
their number indicated by braces.
}
\label{fig:doubl1}
\end{figure}

The set of D7-branes we have introduced has a nice interpretation in the 
IIA side, as we discuss in the following. The D7$_1$-branes (D7$_2$-branes) 
map to lower (upper) half D6$'$-branes ending on the NS branes, as depicted 
in figure~\ref{fig:doubl1}. The spectrum (\ref{spec37odd}) and superpotential
(\ref{superp1}) are nicely reproduced by this IIA brane configuration, 
once we take into account the flavour doubling effect \cite{bhkl}.

Several other features in this map are appealing as well. Gauge symmetries 
on the D7-branes correspond precisely to the gauge groups on the half 
D6$'$-branes (and to global chiral symmetries in the four-dimensional field 
theory \cite{hanbro}). The six-dimensional theory on the half-D6 branes 
contains some bifundamentals $\sum_{i=1}^P (v_P,w_P)$, which are present in 
7$_1$-7$_2$ sector of the IIB construction. Finally, the tadpole conditions 
on the IIB side correspond to conservation of RR charge in the IIA side. 
The constant $C_0$ in (\ref{anom1}) maps to the IIA cosmological constant. 
Also, the transitions in which the number of field theory flavours changes 
are explained as in the models of Section~2. As a final comment, let us 
note that, just like the orbifold models of  Section~2, these IIB 
orientifolds contain 7$_i$-7$_i$ states not present in the IIA configuration. 
It is not clear whether these states are actually massless before taking 
the ALE limit of the Taub-NUT space. Similar comments apply 
to the remaining orientifold models in this section.

\subsection{Even order case, $N=2P$}

\subsubsection{Models with $\Tr\gamma_{\theta^N,3}=+1$}

Let us start with the models studied in section 4.2.1. Recall that there are 
two possibilities for the orientifold group. We consider first the projection 
(\ref{gorientone}), that is ${\IZ_N}+\Omega_1{\IZ_N}$. Let us introduce a 
set of D7$_1$- and D7$_2$-branes, with
{\small
\beqa
\gamma_{\theta,7_1} & = & \diag(e^{\pi i \frac{1}{N}} 1_{w_1}, \ldots 
e^{\pi i \frac{2P-1}{N}} 1_{w_{P}}, e^{\pi i \frac{2P+1}{N}} 1_{w_{P}}, 
\ldots, e^{\pi i \frac{4P-1}{N}} 1_{w_1}) 
\label{D7vecone}
\eeqa
}
and the analogous $\gamma_{\theta,7_2}$ with $w_i$ instead of $v_i$. 
Choosing the $SO$ projection (\ref{soproj}) on the D3-brane forces the 
following projection on D7-branes
{\footnotesize
\beqa
\gamma_{\Omega_1,7_1}  =  \left( \begin{array}{cccccc}
 & & & & & 1_{w_1} \\
 & & & &\cdots &  \\
 & & & 1_{w_P} & & \\
 & & -1_{w_P} & & & \\
 & \cdots  & & & &  \\
 -1_{w_1}& & & & & \\
\end{array}  \right) \quad ; \quad
\gamma_{\Omega_1,7_2}  =  \left( \begin{array}{cccccc}
 & & & & & 1_{v_1} \\
 & & & &\cdots &  \\
 & & & 1_{v_P} & & \\
 & & 1_{v_P} & & & \\
 & \cdots  & & & &  \\
 1_{v_1} & & & & & \\
\end{array}  \right) \quad
\label{D7vectwo}
\eeqa 
}
The resulting 3-7 spectrum is
{\small
\beq
\sum_{i=1}^P \;[\; (n_i,{\ov w_i}) + ({\ov n_{i-1}},w_i)\; ]\; + 
\sum_{i=1}^P \; [\; ({\ov n_i},v_i) + (n_{i-1},{\ov v_i})\;]
\label{spec37vec}
\eeq
}
with a D7-brane group $\prod_{i=1}^P SU(v_i)\times \prod_{i=1}^P SU(w_i)$. 
The superpotential is
\beqa
\sum_{i=0}^{P-1} \;[\; (n_i,{\ov n}_{i+1})(n_{i+1},{\ov w}_{i+1}) 
(w_{i+1},{\ov n}_i) - ({\ov n}_i,n_{i+1})({\ov n}_{i+1},v_{i+1}) 
({\ov v}_{i+1},n_i) \;]\;
\label{sup37vec}
\eeqa

Let us discuss the computation of tadpoles. As explained at the beginning 
of section 4.2, in this type of models (with $\Tr \gamma_{\theta^N,3}=+1$)
 the orientifold action on the $\IZ_2$ twisted sector is such that 
antisymmetric combinations of left and right movers survive in the RR 
closed string 
sector. The orientifold action is such that the two contributions 
$Z(1,\theta^k)$ and $Z(\theta^P,\theta^k)$ to the Klein bottle have negative 
relative sign. Taking these points into account, the tadpoles are
\beqa
\cc & = & \sum_{k=1}^{N-1} (\Tr \gamma_{\theta^k,7_1})^2 + 
\sum_{k=1}^{N-1} (\Tr \gamma_{\theta^k,7_2})^2  - 2\sum_{k=1}^{N-1} 
\Tr \gamma_{\theta^k,7_1} \Tr\gamma_{\theta^k,7_2} \nonumber \\
\cm & = & 
-16 \sum_{\stackrel{\textstyle k=1}{k\neq N/2}}^{N-1}\Tr(\gamma_{\theta^k 
\Omega_1,7_1}^T \gamma_{\theta^k \Omega_1,7_1}^{-1})  -16 
\sum_{\stackrel{\textstyle k=1}{k\neq N/2}}^{N-1} \Tr(\gamma_{\theta^k 
\Omega_1,7_2}^T \gamma_{\theta^k \Omega_1,7_2}^{-1}) \nonumber \\ 
\ck & = & \sum_{\stackrel{\textstyle k=1}{k\neq N/2}}^{N-1} (64 - 64)
\label{zero1}
\eeqa
The relative sign for the Klein bottle mentioned above makes the corresponding 
amplitude vanish. We also see that using the  properties
\beqa
\Tr(\gamma_{\theta^k \Omega_1,7_i}^T \gamma_{\theta^k \Omega_1,7_i}^{-1}) 
=-\Tr(\gamma_{\theta^{k+P} \Omega_1,7_i}^T \gamma_{\theta^{k+P} 
\Omega_1,7_i}^{-1}) 
\eeqa
the contributions from the $k$ and $k+P$ pieces cancel, and the M\"obius 
amplitude vanishes. Thus the tadpole conditions are obtained 
merely from the cylinder piece, leading to the constraints
\beqa
\Tr \gamma_{\theta^k,7_1} +\Tr\gamma_{\theta^k,7_2} = 0 \quad {\rm for} 
\;\; k\neq 0
\label{tadD7again}
\eeqa
They correspond to cancellation of non-abelian anomalies in the D3-brane 
world-volume theory. The most general solution is given by $v_i=w_i+C_0$,
which states that each gauge group has an equal number of fundamental and 
antifundamental multiplets.

\begin{figure}
\centering
\epsfxsize=4in
\hspace*{0in}\vspace*{.2in}
\epsffile{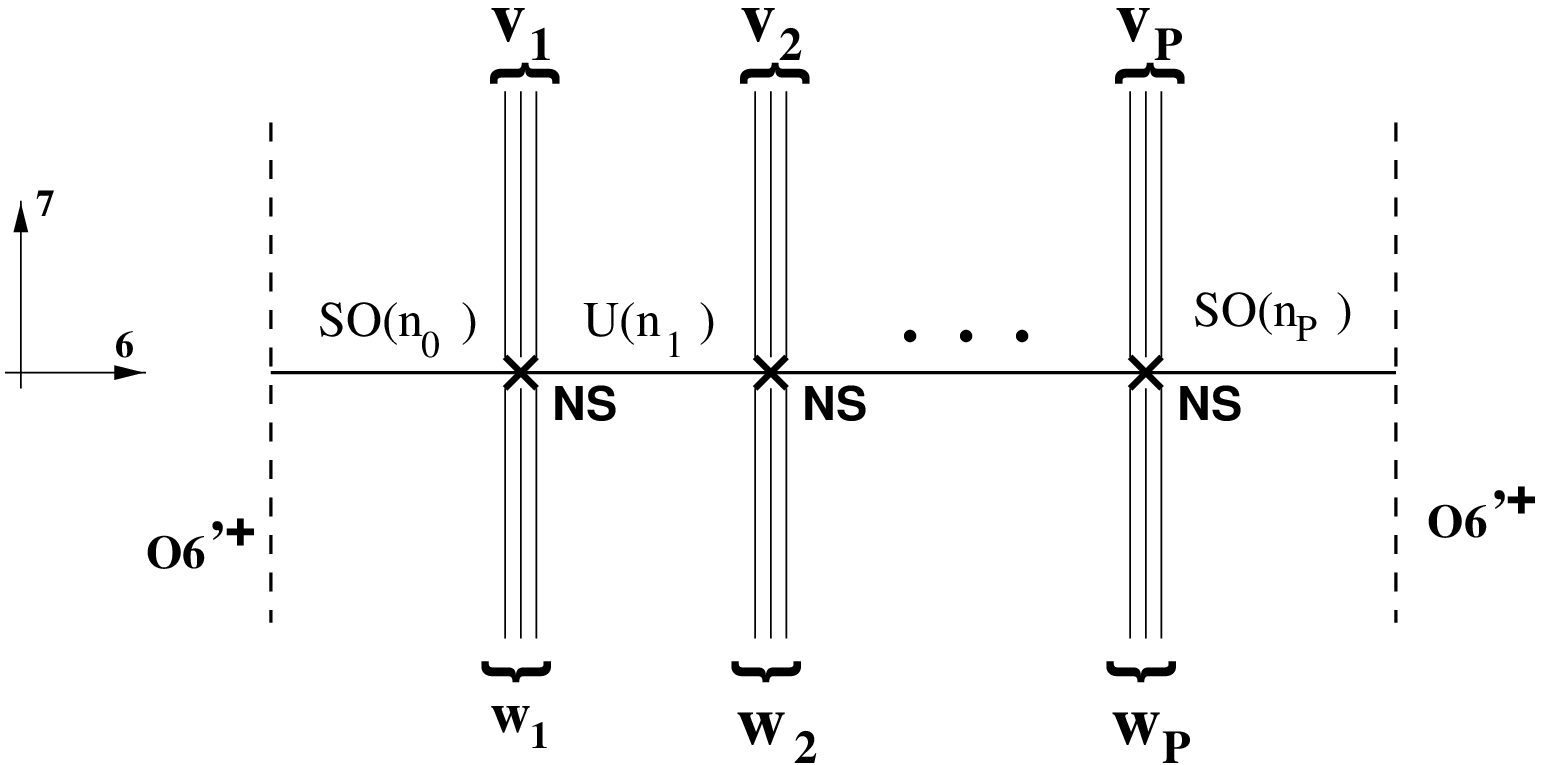}
\caption{\small The IIA brane configuration corresponding to the even 
order IIB orientifold with $\Tr \gamma_{\theta^N,3}=+1$, with a more 
general distribution of D7-branes. 
}
\label{fig:doubl2}
\end{figure}

As in the orientifold models of the previous section, this configuration 
of D7-branes has a direct interpretation in the IIA side. The IIA brane 
configuration is schematically depicted in figure \ref{fig:doubl2}. It 
reproduces the spectrum and superpotential interactions computed in the 
IIB side. As usual, the constant $C_0$ corresponds to the IIA cosmological 
constant.

\medskip

For the orientifold group (\ref{gorientitwo}), that is ${\IZ_N} +\alpha 
\Omega_1{\IZ_N}$, things work analogously. If we introduce a set of D7$_1$- 
and D7$_2$-branes, the corresponding Chan-Paton matrices have the structure 
(\ref{D7vecone}), (\ref{D7vectwo}). The resulting 3-7 spectrum and 
superpotential are given by (\ref{spec37vec}), (\ref{sup37vec}). The tadpoles 
are given by (\ref{zero1}) with the only modifications of replacing the 
matrices $\gamma_{\theta^k\Omega_1,7_i}$ by $\gamma_{\theta^k\alpha
\Omega_1,7_i}$ in $\cm$, and letting $k$ run through the complete range 
$0\ldots N-1$ in $\cm$ and $\ck$. In any event, the vanishing of 
the Klein bottle and M\"obius strip amplitudes also follows in this case. The 
only tadpoles are generated from the cylinder amplitudes, and give the same 
constraint (\ref{tadD7again}) on D7-brane matrices. These conditions ensure 
the cancellation of gauge anomalies on the D3-brane world-volume. The type 
IIA configuration is given by figure \ref{fig:doubl2} if one flips the sign 
of the left O6$'$-plane.

\subsubsection{Models with $\Tr\gamma_{\theta^N,3}=-1$}

Here we consider the models constructed in section 4.2.2. Recall that the 
orientifold group had two possible structures. We start by studying 
(\ref{gorone}), {\em i.e.} ${\IZ_N}+\Omega_1{\IZ_N}$. A quite general 
configuration is obtained by introducing a set of D7$_1$- and 
D7$_2$-branes with
{\small
\beqa
\gamma_{\theta,7_1} & = & \diag(1_{w_0}, e^{2\pi i \frac{1}{N}} 1_{w_1}, 
\ldots, e^{2\pi i \frac{P-1}{N}} 1_{w_{P-1}},e^{2\pi i \frac{P}{N}} 
1_{w_P}, e^{2\pi i\frac{P+1}{N}} 1_{w_{P-1}}, \ldots, e^{2\pi 
i\frac{2P-1}{N}} 1_{w_1}) \quad \quad \quad
\label{D7novecone}
\eeqa
}
(and the analogous $\gamma_{\theta,7_2}$ with $v_i$ instead of $w_i$) and
\beqa
\gamma_{\Omega_1,7_1}  = {\scriptsize \left( \begin{array}{cccccc}
\varepsilon_{w_0} & & & & & \\
  & & & & & 1_{w_1} \\
  & & & & 1_{w_{P-1}}  & \\
  & & & \varepsilon_{w_P}  & & \\
  & & -1_{w_{P-1}}  & & & \\ 
  & -1_{w_1}   & & & & \\
\end{array}  \right) } \;\; ; \;\;
\gamma_{\Omega_1,7_2}  = {\scriptsize \left( \begin{array}{cccccc}
1_{v_0} & & & & & \\
  & & & & & 1_{v_1} \\
 & & & & 1_{v_{P-1}} & \\
 & & &  1_{v_P}  & & \\
 & &  1_{v_{P-1}} & & & \\
 & 1_{v_1} & & & & \\
\end{array}  \right) } \quad
\label{D7novectwo}
\eeqa 
The resulting 3-7 spectrum is given by
\begin{eqnarray}
(n_1,w_0) + \sum_{i=1}^{P-1} \;[\; (n_{i+1},{\ov w_i}) + ({\ov n_{i}},w_i) 
\;]\; + ({\ov n_P},w_P) + \nonumber \\ 
+({\ov n_1},v_0) + \sum_{i=1}^{P-1} \;[\; ({\ov n_{i+1},v_i) 
+ (n_{i}},{\ov v_i})\;]\; + (n_P,v_P) 
\label{spec37novec}
\end{eqnarray}
where the D7-brane groups are $USp(w_0)\times \prod_{i=1}^{P-1} SU(w_i) 
\times USp(w_P)$ and $SO(v_0)\times \prod_{i=1}^{P-1} SU(v_i)\times SO(v_P)$. 
The superpotential is
\beqa
W= & \bYasymm_1(n_1,w_0)^2 + \sum_{i=1}^{P-1} (n_i,{\ov n}_{i+1}) 
(n_{i+1},{\ov w}_i) (w_i,{\ov n}_i) + \Yasymm_P({\ov n}_P,w_P)^2 + \nonumber\\ 
+ & \Ysymm_1({\ov n}_1,v_0)^2 + \sum_{i=1}^{P-1} ({\ov n}_i,n_{i+1}) 
({\ov n}_{i+1},v_i) ({\ov v}_i,n_i) + {\ov {\Ysymm}}_P (n_P,v_P)^2 
\label{sup37novec}
\eeqa

\begin{figure} 
\centering 
\epsfxsize=4in
\hspace*{0in}
\vspace*{.2in} 
\epsffile{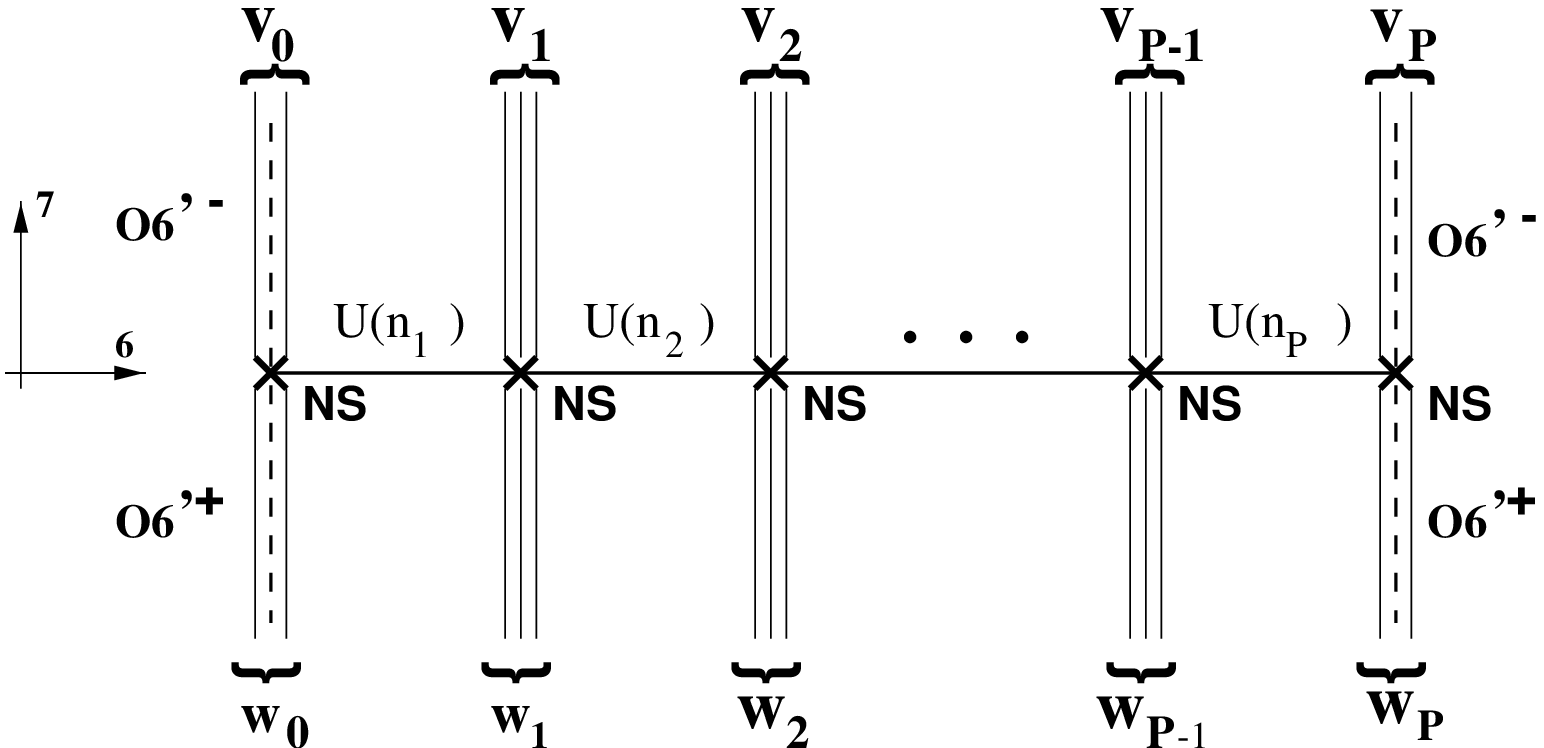} 
\caption{\small The IIA
brane configuration corresponding to the even order IIB orientifold
with $\Tr \gamma_{\theta^N,3}=-1$, with a more general distribution of 
D7-branes.}
\label{fig:doubl3} 
\end{figure}

Let us sketch the computation of tadpoles. In this case the orientifold 
projection on the closed string sector chooses RR symmetric combinations. 
Consequently, the relative sign between the two contribution to the Klein 
bottle is positive. The corresponding tadpoles are 
\beqa 
\cc & = & 
\sum_{k=1}^{N-1} (\Tr\gamma_{\theta^k,7_1})^2 + \sum_{k=1}^{N-1} 
(\Tr\gamma_{\theta^k,7_2})^2 - 2\sum_{k=1}^{N-1} \Tr\gamma_{\theta^k,7_1} 
\Tr\gamma_{\theta^k,7_2} \nonumber \\ 
\cm & = & -16 \sum_{k=1}^{N-1} 
\Tr(\gamma_{\theta^k \Omega_1,7_1}^T \gamma_{\theta^k \Omega_1,7_1}^{-1})  
-16 \sum_{k=1}^{N-1} 
\Tr(\gamma_{\theta^k\Omega_1,7_2}^T \gamma_{\theta^k \Omega_1,7_2}^{-1}) 
\nonumber \\ 
\ck & = & \sum_{\stackrel{k=1}{k\neq N/2}}^{N-1} (64 + 64)  
\label{nonzero1} 
\eeqa 
In this case, the matrices have the properties 
\beqa 
\Tr(\gamma_{\theta^k \Omega_1,7_i}^T\gamma_{\theta^k \Omega_1,7_i}^{-1}) 
=\Tr(\gamma_{\theta^{k+P}\Omega_1,7_i}^T 
\gamma_{\theta^{k+P} \Omega_1,7_i}^{-1}) = \pm
\Tr\gamma_{\theta^{2k},7_i} 
\eeqa 
with the negative (positive) sign for D7$_1$ (D7$_2$) branes. They can be used 
to rewrite the tadpoles as 
\beqa
 \sum_{k=1}^{N/2} & (\Tr \gamma_{\theta^{2k-1},7_1} -  
\Tr\gamma_{\theta^{2k-1},7_2})^2 + & \nonumber \\ 
+ \sum_{k=1}^{N/2-1}
& [\; (\Tr \gamma_{\theta^{2k},7_1}-\Tr\gamma_{\theta^{2k},7_2})^2 + & 
32 (\Tr \gamma_{\theta^{2k},7_1} - \Tr\gamma_{\theta^{2k},7_2}) + 256 \;] 
\eeqa 
The factorization leads to the constraints
\beq
\Tr\gamma_{\theta^k,7_1} -\Tr\gamma_{\theta^k,7_2} + 16 \delta_{k,0\;{\rm
mod}\; 2} = 0 \quad {\rm for} \; k\neq 0 
\label{tadnovecone} 
\eeq 
Their general solution, $w_i=v_i+C_0-8\delta_{i,0}-8 \delta_{i,P}$, 
ensures the cancellation of gauge anomalies on the D3-brane. The corresponding 
type IIA configuration, reproducing the spectrum (\ref{spec37novec}) and
interactions (\ref{sup37novec}), is shown in figure~\ref{fig:doubl3}. 

\medskip

We now consider the case of the orientifold group (\ref{gortwo}) 
$\IZ_N+\alpha\Omega_1\IZ_N$. A quite general choice for D7-brane Chan-Paton 
matrices is given by the (\ref{D7novecone}) for the orbifold twist, and
{\small
\beqa
& \gamma_{\alpha\Omega_1,7_1}  = & \left( \begin{array}{cccccc}
1_{w_0} & & & & &  \\
 & & & & & \alpha 1_{w_1} \\
 & & & & \alpha^{P-1} 1_{w_{P-1}} & \\
 & & & \alpha^P \varepsilon_{w_P} & & \\
 & & \alpha^{-(P-1)} 1_{w_{P-1}} & & & \\ 
 & \alpha^{-1}1_{w_1} & & & & \\
\end{array}  \right) 
\nonumber \\
& \gamma_{\Omega_1,7_2}  =  & \left( \begin{array}{cccccc}
\alpha^P \varepsilon_{v_0} & & & & & \\
 & & & & & \alpha^{-(P-1)}1_{v_1} \\
 & & & & \alpha^{-1} 1_{v_{P-1}} & \\
 & & & 1_{v_P} & & \\
 & & \alpha 1_{v_{P-1}} & & & \\
 & \alpha^{P-1} 1_{v_1} & & & & \\
\end{array}  \right) 
\label{cpnovectwo}
\eeqa 
}
for the orientifold action. The resulting 3-7 spectrum and superpotential 
coincide with (\ref{spec37novec}), (\ref{sup37novec}), but the D7-brane 
symmetries are $SO(w_0)$$\times \prod_{i=1}^{P-1} SU(w_i)$$\times USp(w_P)$ 
and $USp(v_0)$$\times \prod_{i=1}^{P-1} SU(v_i)$$\times SO(v_P)$. 

The tadpoles are given by (\ref{nonzero1}) with the replacement of 
$\Tr\gamma_{\theta^k\Omega_1}$ by 
$\Tr\gamma_{\theta^k\alpha\Omega_1,7_i}$, and summing over the full range 
$k=0,\ldots,N-1$ in $\ck$ and $\cm$. 
The matrices satisfy the property 
\beqa 
\Tr(\gamma_{\theta^k \alpha\Omega_1,7_i}^T 
\gamma_{\theta^k\alpha \Omega_1,7_i}^{-1}) =
\Tr(\gamma_{\theta^{k+P} \alpha\Omega_1,7_i}^T
\gamma_{\theta^{k+P} \alpha \Omega_1,7_i}^{-1}) 
= \pm \Tr\gamma_{\theta^{2k+1},7_i} 
\eeqa 
now with the positive (negative) sign for D7$_1$ (D7$_2$) branes. Using 
this relation, the tadpoles can be written as 
\beqa  
\sum_{k=1}^{N/2-1} & [ (\Tr \gamma_{\theta^{2k},7_1}-
\Tr\gamma_{\theta^{2k},7_2})^2 + & \nonumber \\ 
+ \sum_{k=1}^{N/2}
& [\; (\Tr\gamma_{\theta^{2k-1},7_1} - \Tr\gamma_{\theta^{2k-1},7_2})^2 &
- 32 (\Tr\gamma_{\theta^{2k-1},7_1} - \Tr\gamma_{\theta^{2k-1},7_2})+256\; ] 
\eeqa
The factorization of this expression yields the constraints
\beqa 
\Tr \gamma_{\theta^{k},7_1} - \Tr\gamma_{\theta^{k},7_2} - 16
\delta_{k,1\,{\rm mod}\, 2} =0 \quad {\rm for} \; k\neq 0 
\label{tadnovectwo}
\eeqa 
Their general solution is $w_i=v_i + 8\,\delta_{i,0} - 8\,\delta_{i,P+1} + 
C_0$, which ensures the cancellation of anomalies on the D3-brane probe.
The corresponding type IIA configuration is that obtained from 
figure~\ref{fig:doubl3} by inverting the orientation of the fork on the left. 

\medskip

\section{Conclusions}

In this paper we have analyzed several features of T-duality for some type 
IIA brane configurations preserving four supercharges. These 
configurations realize four-dimensional $\NN=1$ supersymmetric field 
theories with interesting chirality properties. The T-dual configurations 
have been found to be $\NN=1$ orientifolds of $\IC^2/\IZ_N$ singularities, 
which we have explicitly constructed and matched with the IIA models.

The discussion has enlightened several aspects about the string theory 
embedding of these field theories. We have shown that several seemingly 
exotic effects in the type IIA construction have a perfectly smooth and 
standard realization when translated to the type IIB picture. A crucial 
fact for this feature is the peculiar mapping of spacetime directions in 
this T-duality. Specifically, the positive and negative regions of the 
IIA coordinate $x^7$ are mapped to two different complex planes in 
$\IC^2/\IZ_N$ in the IIB model. We have provided extensive evidence for 
this fact, mainly based on a precise matching of the different type IIA 
and IIB models.

We also would like to stress that the type IIB realization of these field 
theories is very similar to that of other chiral gauge theories. This 
points towards a more unified description of the string theory 
configurations yielding chiral field theories.

Finally, even though we have centered on understanding the properties of 
the string theory configurations, the type IIB picture of D3-brane probes 
that we have constructed also presents some advantages for the study of 
the large N limit of the field theory. It would be interesting to find out 
the relation between these models and other chiral gauge theories from the 
point of view of the AdS/CFT correspondence.

\bigskip

\begin{center}
{\bf Acknowledgements}
\end{center}

It is our pleasure to thank J.~Erlich, A.~Hanany, L.~E.~Ib\'a\~nez, 
B.~Janssen, A.~Kapustin, A.~Karch, P.~Meessen and A.~Naqvi for useful 
conversations. A.~M.~U. is grateful to  G.~Aldazabal and D.~Badagnani for 
their insights about orientifold constructions, and also to M.~Gonz\'alez 
for kind encouragement and support, and to the Center for Theoretical Physics 
at M.~I.~T. for hospitality. The research of J.~P. is supported by the 
US. Department of Energy under Grant No. DE-FG02-90-ER40542. The research 
of R.~R. is supported by the Ministerio de Educaci\'on y Cultura (Spain) 
under a FPU Grant. The research of A.~M.~U. is supported by the Ram\'on 
Areces Foundation (Spain).

\end{document}